\documentclass[12pt,preprint]{aastex}
\usepackage{natbib}
\usepackage{amsmath}
\usepackage{amsbsy}
\usepackage{color}
\usepackage{rotating}
\usepackage[breaklinks,colorlinks,urlcolor=green,citecolor=blue,linkcolor=red]{hyperref}

\defcitealias{Lynch2013}{L\&E13}
\slugcomment{{\today}}

\shorttitle{Reconnection in CME Current Sheets}
\shortauthors{Lynch et al.}

\begin{document}


\title{Reconnection Properties of Large-Scale Current Sheets During
\\ Coronal Mass Ejection Eruptions}


\author{B. J. Lynch\altaffilmark{1}, J. K. Edmondson\altaffilmark{2},
M. D. Kazachenko\altaffilmark{1}, and S. E. Guidoni\altaffilmark{3}}

%

%
\affil{\altaffilmark{1}Space Sciences Laboratory, University of
California, Berkeley, CA 94720, USA}
\affil{\altaffilmark{2}Climate and Space Sciences and Engineering
Department, University of Michigan, Ann Arbor, MI, 48109, USA}
\affil{\altaffilmark{3}Heliophysics Science Division, NASA Goddard
Space Flight Center, Greenbelt, MD, 20771, USA}

\begin{abstract}

We present a detailed analysis of the properties of magnetic
reconnection at large-scale current sheets in a high cadence version
of the \citet{Lynch2013} 2.5D MHD simulation of sympathetic magnetic
breakout eruptions from a pseudostreamer source region.
We examine the resistive tearing and breakup of the three main
current sheets into chains of X- and O-type null points and follow
the dynamics of magnetic island growth, their merging, transit, and
ejection with the reconnection exhaust.
For each current sheet, we quantify the evolution of the length-to-width
aspect ratio (up to $\sim$100:1), Lundquist number ($\sim$10$^3$),
and reconnection rate (inflow-to-outflow ratios reaching $\sim$0.40).
We examine the statistical and spectral properties of the fluctuations
in the current sheets resulting from the plasmoid instability,
including the distribution of magnetic island area, mass, and flux
content.
We show that the temporal evolution of the spectral index of the
reconnection-generated magnetic energy density fluctuations appear
to reflect global properties of the current sheet evolution.
Our results are in excellent agreement with recent, high resolution
reconnection-in-a-box simulations even though our current sheets'
formation, growth, and dynamics are intrinsically coupled to the
global evolution of sequential sympathetic CME eruptions.

\end{abstract}

\keywords{magnetic reconnection --- magnetohydrodynamics (MHD) ---
current sheets --- Sun: corona --- Sun: coronal mass ejections
(CMEs) --- Sun: flares}

%
%
\section{Introduction}
\label{s:intro}

The magnetic reconnection rate from the Sweet-Parker reconnection
model \citep{Sweet1958,Parker1963} for elongated current sheets is
generally too ``slow'' to account for the observed rapid flux
transfer in solar flares.
The solar corona is extremely conductive so the magnetic Reynolds
number (the Lundquist number) is of the order of $S=4\pi\sigma V_A
L / c^2 \sim 10^{10-12}$ where $V_A$ is the Alfven speed, $L$ is a
characteristic global length scale, $\sigma$ is the plasma conductivity,
and $c$ is the speed of light.
The scaling of the Sweet-Parker reconnection inflow velocity $V_{in}$
to the reconnection outflow velocity (taken as $V_A$) goes as
$V_{in}/V_A \sim S^{-1/2}$, yielding theoretical inflow speeds of
$V_{in} \lesssim 10^{-5-6} V_A$.
The \citet{Petschek1964} reconnection model is able to obtain the
much greater inflow speeds needed for ``fast" reconnection, e.g.
$V_{in} \sim 0.10 V_A$, but requires a much shorter current sheet
(essentially a single X-point) and a very localized enhancement of
the magnetic resistivity (equivalently, a very localized depletion
of the plasma conductivity).
In collisionless reconnection models, the scales of kinetic dissipation
effects are sufficiently small ($\sim$$10^2$~cm) that these models
are also able to produce fast, Petschek-like reconnection scenarios.
Given that the highest resolution observations of current sheets
in the corona are still macroscopic in scale (i.e., $10^{6-7}$~cm),
it is not clear \emph{if} or \emph{how} the microscopic -- and
presently unobservable -- Petschek or Petschek-like reconnection
processes relate to these large-scale observations.
Therefore, a tremendous body of work on detailed reconnection
studies, both analytical and numerical, has been dedicated to
attempting to resolve this situation \citep[e.g., see][and references
therein]{Cassak2013a}.
Significant progress has been made in understanding the physical
processes that effectively ``speed up" the
reconnection rates associated with large-scale current sheets.

One of the ways to speed up the reconnection is the onset and
development of instabilities that result in current sheet tearing,
breakup, and the formation of magnetic island plasmoids
\citep[][]{Furth1963,Forbes1983,Biskamp1996}. The onset of the
resistive tearing mode has been characterized by the fastest growing
wavelengths associated with linearized perturbation analysis and
advances in numerical modeling have enabled simulations of the
highly nonlinear time-dependent evolution of the plasmoid instability
\citep[e.g.,][and references
therein]{Forbes1991b,Karpen1998,Karpen2012,Loureiro2007,Lin2009,Samtaney2009,Bhattacharjee2009,Huang2010,Ni2010}.

An alternative way to speed up reconnection is to make the current
sheet thin enough that traditional (resistive) MHD formalism breaks
down.
If the current sheet dissipation region becomes smaller than the
ion skin depth then the system enters a collisionless regime which
can generate reconnection rates orders of magnitude faster than
Sweet-Parker. Modeling this regime is accomplished by either including
the generalized Ohm's law (``Hall MHD") or by going to a hybrid
(electron fluid, ion particle) or fully kinetic particle treatment;
see, e.g., the results of the GEM reconnection challenge
\citep[][]{Birn2001a,Birn2001b,Hesse2001,Kuznetsova2001,Ma2001,Otto2001,Pritchett2001,Shay2001}.

In fact, these two approaches to speed up the reconnection -- current
sheet tearing and plasmoid formation in traditional MHD and the
inclusion of more physics and particle effects in the numerical
modeling (Hall MHD, hybrid, particle codes) -- are not mutually
exclusive. The break-up of large-scale current sheets into chains
of magnetic islands appears to be a robust and universal feature
in all of the different types of 2D reconnection modeling
\citep[e.g.,][]{Drake2006b,Daughton2006,Daughton2009,Cassak2009b}.
The discussion by \citet{Edmondson2010} provides some important
physical insight into why the plasmoid formation appears to be such
a robust, universal process.

The physical properties that govern the reconnection dynamics in a
current sheet are determined by the global system. The geometry of
the current sheet, its length and width, are related to the global
scale of the magnetic configuration and the scale at which the
frozen-in flux condition breaks (the diffusion scale), respectively.
In MHD simulations, the diffusion scale is set by the resistivity
model which, at its smallest value, is essentially the scale of the
numerical grid. In simulations that include particle effects, the
diffusion scale is given by the ion skin depth or gyroradius. The
inflow and outflow velocities transporting magnetic flux into the
current sheet and reconnected flux out in the exhaust are likewise
determined by the global system. The inflow velocities are typically
boundary conditions in dedicated reconnection simulations or, as
in our case here, determined by the global dynamics of the magnetic
field, and the outflow velocities are of the order of the Alfven
speed. Since the global system has determined all the key parameters
governing the reconnecting current sheet, the additional constraints
the system must operate under, such as the conservation of mass and
magnetic flux, means the system is actually \textit{over-determined}.
The break-up of the current sheet and the formation of magnetic
islands resolves this over-determination by introducing new scales
into the system that allow both the conservation laws and the global
constraints on the current sheet reconnection properties to be met
simultaneously.

The purpose of this paper is to examine, in detail, the properties
and evolution of magnetic reconnection and the magnetic island
plasmoids generated in the current sheets that arise in the sympathetic
eruption scenario of \citet{Lynch2013}, hereafter abbreviated as
\citetalias{Lynch2013}.
The paper is structured as follows. 
In Section~\ref{s:overview}, we briefly discuss the MHD code and
review the \citetalias{Lynch2013} simulation results that
self-consistently create the three large-scale current sheets during
two sequential CME eruptions.
In Section~\ref{s:simdiff}, we compare the global properties and
evolution of the current sheets in terms of their Lundquist number,
inflow and outflow properties, and reconnected magnetic flux.
In Section~\ref{s:islands}, we present distribution functions of
the island area, mass and flux content, and examine the spectral
properties of the magnetic fluctuations in the current sheets.
In Section~\ref{s:discussion}, we discuss the implications of our
results and the direction of future work.

%
%
\section{Numerical Simulation Methods and Summary of Previous Results}
\label{s:overview}

\subsection{\textit{ARMS}: Adaptively Refined MHD Solver}
\label{ss:arms}

The Adaptively Refined MHD Solver \citep[\textit{ARMS};][]{DeVore2008}
calculates solutions to the 3D nonlinear, time-dependent MHD equations
using a finite-volume flux-corrected transport numerical scheme
\citep{DeVore1991}.
\textit{ARMS} is fully integrated with the adaptive mesh toolkit
PARAMESH \citep{MacNeice2000} to handle solution-adaptive grid
refinement and support efficient multi-processor parallelization.
\textit{ARMS} has been used to perform a wide variety of numerical
simulations of dynamic phenomena in the solar atmosphere, including
3D magnetic breakout CME initiation \citep[][]{DeVore2008,Lynch2008},
the eruption of coronal jets \citep{Pariat2009,Pariat2010}, the
interaction between closed and open fields at streamer belt boundaries
\citep{Edmondson2009} and during CME eruptions \citep{Masson2013},
and the detailed examination of current sheet formation, magnetic
reconnection, and magnetic island creation
\citep{Edmondson2010,Karpen2012,Guidoni2016}.

For the simulation discussed herein, we use \textit{ARMS} to solve
the ideal MHD equations in Cartesian coordinates,
\begin{equation}
    \frac{\partial \rho}{\partial t} + \nabla \cdot \left( \rho
    \mathbf{V} \right) = 0 ,
    \label{e1}
\end{equation}
\begin{equation}
    \frac{\partial\left( \rho \mathbf{V} \right)}{\partial t}  + \nabla
    \cdot \left( \rho \mathbf{V} \mathbf{V} \right) + \nabla P =
    \frac{1}{4\pi}\left( \nabla \times \mathbf{B} \right) \times \mathbf{B} ,
    \label{e2}
\end{equation}
\begin{equation}
    \frac{\partial T}{\partial t} + \nabla \cdot \left( T \mathbf{V}
    \right) + \left( \gamma-1 \right) T \left( \nabla \cdot \mathbf{V}
    \right) = 0 ,
    \label{e3}
\end{equation}
\begin{equation}
    \frac{\partial \mathbf{B}}{\partial t} = \nabla \times \left(
    \mathbf{V} \times \mathbf{B} \right).
    \label{e4}
\end{equation}

\noindent The variables retain their usual meaning: mass density
$\rho$, velocity $\mathbf{V}$, magnetic field $\mathbf{B}$, and we
have written the energy equation in terms of the plasma temperature
$T$. The ratio of specific heats is $\gamma=5/3$ and the ideal gas
law $P=2(\rho/m_p)k_BT$ closes the system.
Additionally, while there is no explicit magnetic resistivity in
the equations of ideal MHD, necessary and stabilizing numerical
diffusion terms introduce an effective resistivity on very small
spatial scales, i.e. the size of the grid. In this way, magnetic
reconnection can occur when sharp magnetic gradients
of field component reversals and their associated current sheet
features have been  compressed to the local grid resolution scale.

The full computational domain is $x \in [-5, 5]$, $y \in [1, 21]$
in units of the characteristic length scale $L_0 = 10^9$~cm. There
are six total levels of static grid refinement that vary in the $y$
direction.
For the $y \in [1,11]$ region analyzed herein: $1 \le y \le 7.094$
is level six, $7.094 < y \le 9.750$ is level five, and $9.750 < y
\le 11$ is level four. The highest refinement region corresponds
to an effective $1024 \times 1024$ resolution and we have interpolated
the lower refinement regions to this resolution. The current sheets
that we will examine remain entirely in the level six portion of
the domain.

The initial magnetic field configuration is constructed from the
magnetic vector potential of a series of line dipoles to create the
pseudostreamer arcades embedded in a uniform vertical background
field \citep[see also][]{Edmondson2010}. The background field
strength is $B_0 = 5$~G whereas the line dipoles yield field strengths
in the pseudostreamer arcades of $\sim$35~G. The initial uniform
mass density $\rho_0 = 10^{-16}$~gm~cm$^{-3}$ and pressure $P_0 =
0.01$~dyn~cm$^{-2}$ result in a global plasma beta of $\beta_0 \sim
0.01$ and global Alfven speed $V_{A0} \sim 1400$~km~s$^{-1}$.

The system is energized with shear flows at the lower boundary
parallel to the pseudostreamer arcade polarity inversion lines that
are smoothly ramped up, remain uniform for $\sim$1000~s, and then
are smoothly ramped back down to zero \citepalias[see Figure 1
of][]{Lynch2013}. In order for reconnection to proceed in our system,
the initial symmetry is broken by ramping down the shearing flows
in the left psuedostreamer arcade first and continuing the uniform
shearing in the right arcade for an additional 150~s. This has the
effect of distorting the pseudostreamer X-point by the separation
of the inner and outer spine lines as in the
\citet{Syrovatskii1971,Syrovatskii1978a,Syrovatskii1978b,Syrovatskii1981}
scenario and forming the initial overlying ``magnetic breakout"
current sheet. The development of this current sheet for $t \gtrsim
1250$~s, as well as the subsequent sympathetic CMEs that each form
their own eruptive flare current sheets, is entirely due to the
global response of the system to the accumulated free magnetic
energy supplied by the shearing flows.

\begin{figure*}
\center \includegraphics[width=6.5in]{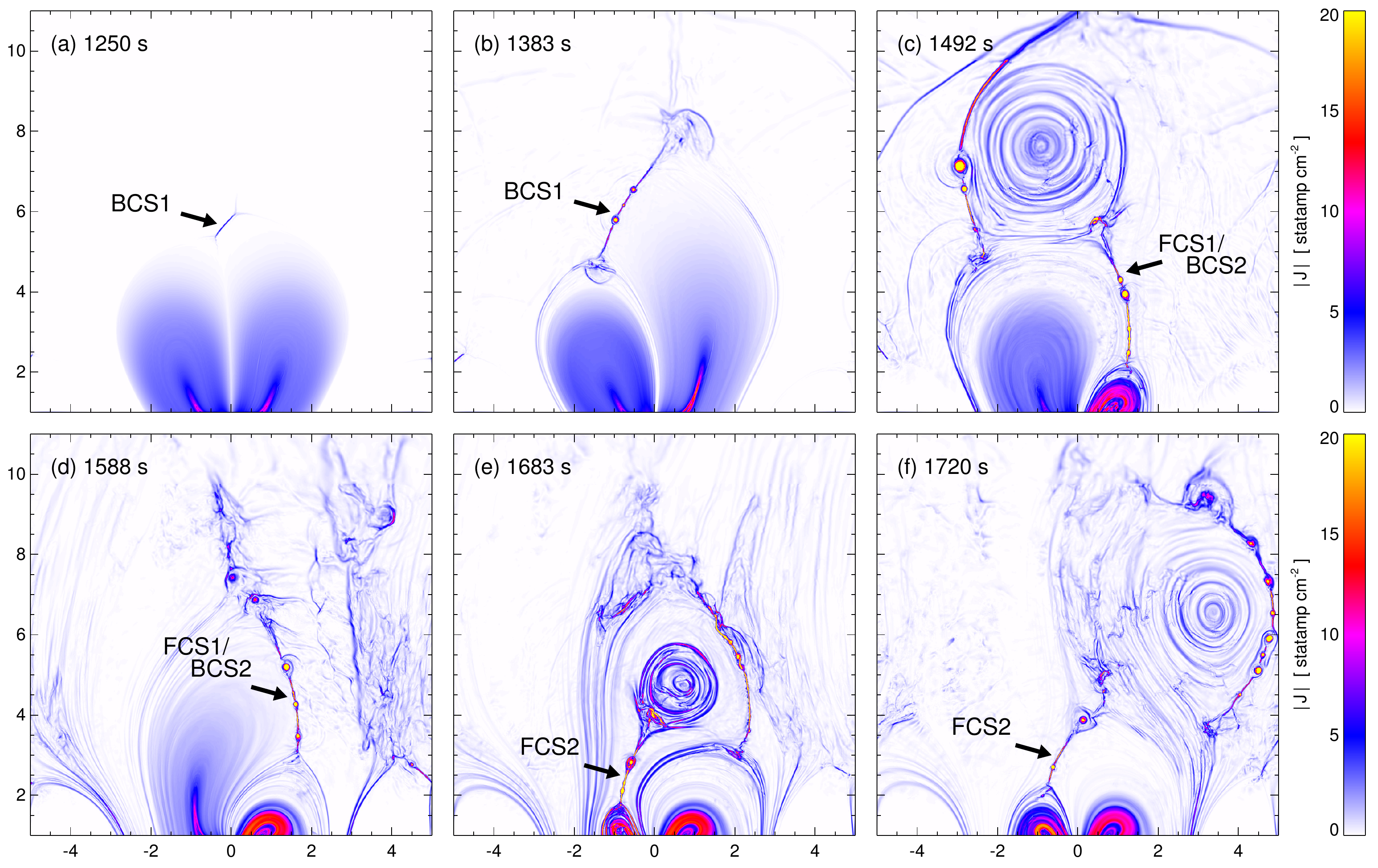} 
\caption{Summary plot of the current density magnitude $|J|$
illustrating the sympathetic magnetic breakout eruption scenario.
Panels (a) and (b) show the breakout phase and BCS1. Panels (c) and
(d) show the first eruption, FCS1/BCS2, and the transition of
FCS1/BCS2 into the breakout phase for the second eruption. Panels
(e) and (f) show the second eruption and FCS2. An animation of this
figure is available as an electronic supplement to the online version
({\tt FIGURE1\_jm.mp4}).  \label{f1}
}
\end{figure*}
%
%

\subsection{Sympathetic Breakout Eruption Scenario}
\label{ss:2013paper}

\begin{figure}
\center \includegraphics[width=3.25in]{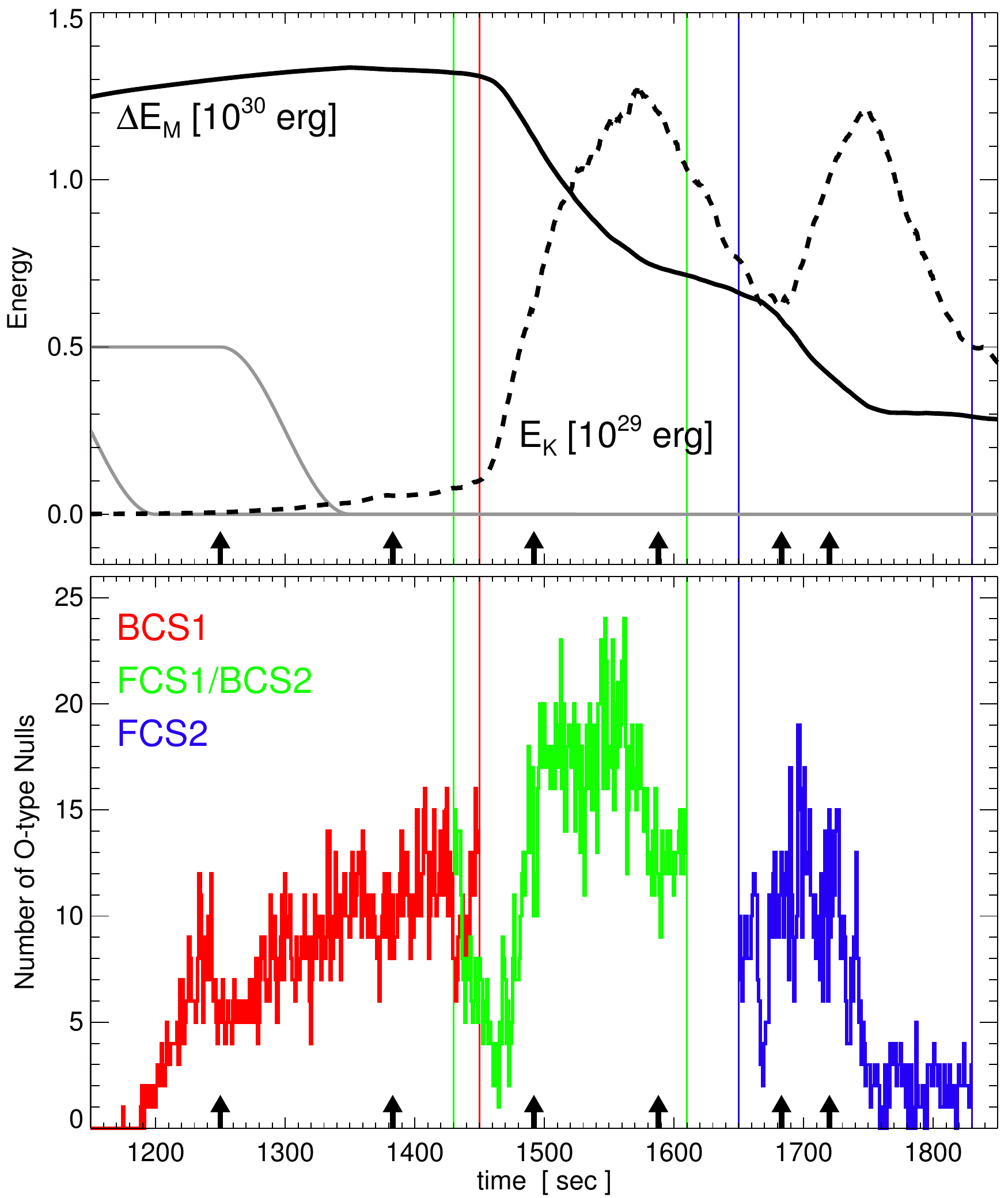}
\caption{Upper panel: global magnetic free energy ($\Delta E_M$)
and kinetic energy ($E_K$) evolution. Lower panel: number of O-type
null points (magnetic islands) in each current sheet (BCS1 red;
FCS1 green; FCS2 blue). The black arrows correspond to the times
shown in the six panels of Figure~\ref{f1}.
\label{f2}
}
\end{figure}
%

\citetalias{Lynch2013} discussed the global evolution and interaction
of the pseudostreamer and background flux systems during the
sympathetic magnetic breakout eruption process.
This was largely an extension of the idea presented by \citet{Torok2011}
who showed that quasi-stable flux ropes anchored in the arcades of
a pseudostreamer could be made to erupt in sequence if a sufficiently
large perturbation was introduced. \citet{Torok2011} highlighted
the evolution of various current sheets that form in response to
this perturbation (which in their numerical simulation was the
eruption of a third flux rope in the vicinity of the pseudostreamer).
Those authors also noted the similarities of the system's magnetic
topology to that required for the magnetic breakout CME initiation
model \citep{Antiochos1999,MacNeice2004,DeVore2008,Karpen2012}.

\citetalias{Lynch2013} confirmed that the sympathetic eruption
sequence could, in fact, be initiated via magnetic breakout and
showed that each eruption resulted in a fast $\gtrsim 1500$~km~s$^{-1}$
flux rope CME that was formed by the flare reconnection above the
polarity inversion line of one of the pseudostreamer arcades.
The simulation resolution was sufficiently high such that we were
able to model the detailed structure and evolution of the current
sheet during the onset and development of the plasmoid instability.
We presented those results qualitatively in the context of the
adaptive mesh refinement runs by \citet{Karpen2012} which showed
that the  onset of ``fast" reconnection associated with the current
sheet tearing and ubiquitous magnetic island formation was directly
responsible for, and essentially defined, the acceleration phase
of the CME eruption.

Figure~\ref{f1} plots six frames of current density magnitude $|J|$
during the sympathetic magnetic breakout eruption scenario.
Panels~\ref{f1}(a), (b) show the period of overlying breakout
reconnection that evolves (relatively) slowly and acts to remove
restraining flux from above the sheared field core which will
eventually become the center of the first erupting flux rope-like
structure. We have labeled this first breakout current sheet BCS1.
Panel~\ref{f1}(c) shows the eruption of the first CME from the right
arcade of the pseudostreamer. The runaway expansion from the
expansion-breakout reconnection positive feedback enables the
formation of the second, essentially vertical current sheet underneath
the rising sheared field core, just as in the standard CSHKP eruptive
flare picture \citep{Carmichael1964, Sturrock1966, Hirayama1974,
Kopp1976}. Here we label the first flare current sheet FCS1/BCS2.
Panel~\ref{f1}(d) shows the system's further evolution where the
continued reconnection at FCS1/BCS2 acts as the overlying breakout
reconnection for the left pseudostreamer arcade, facilitating a
second runaway expansion-breakout reconnection feedback loop.
Finally, in panels~\ref{f1}(e), (f) we show the second CME with its
eruptive flare current sheet labeled FCS2. The accompanying online
animation {\tt FIGURE1\_jm.mp4} shows the complete temporal evolution
of the sympathetic eruptions.

In order to analyze the fine-scale structure and dynamics of the
current sheets, we re-ran the \citetalias{Lynch2013} simulation
with a factor of 10 higher cadence for the output data files. The
\citetalias{Lynch2013} simulation was run on the UCB SSL cluster
``{Shodan}'' (Intel Xeon \emph{Harpertown} architecture, Open MPI
1.4.5 and Intel 12.1 compilers) while the present simulation was
performed on NASA NCCS ``{Discover}'' cluster (Intel Xeon \emph{Sandy
Bridge} architecture, Open MPI 1.7.2 and Portland Group 13.6
compilers).
Despite small, quantitative numerical differences accumulating over
hundreds of thousands of computational time steps, the sympathetic
eruptive flare and CME onset times agree to within $\sim$4\% between
the two simulations (i.e., $\Delta T_{\rm CME1}/T_{\rm CME1} =
50/1470 = 0.034$ and $\Delta T_{\rm CME2}/T_{\rm CME2} = 70/1670 =
0.042$).

The top panel of Figure~\ref{f2} shows the global magnetic and
kinetic energy evolution in our system once the symmetry has been
broken by the  energization flows applied to the lower boundary.
The solid line is the total free magnetic energy $\Delta E_M(t) =
E_M(t) - E_M(0)$ where the initial magnetic energy of the potential
field at $t=0$ is $E_M(0) = 9.96 \times 10^{29}$~erg. The dashed
line is the total kinetic energy $E_K(t)$.  The gray lines indicate
the temporal duration of the boundary shearing flows: by 1150~s the
left arcade driving flow is half of the way through its ramp down,
and the right arcade driving flows begin to be ramped down at 1250~s.
The kinetic energy slowly rises during the period of breakout
reconnection to $\sim$10$^{28}$~erg before the onset of flare
reconnection in FCS1/BCS2 starts the impulsive acceleration of the
first CME -- signaled by the rapid rise of $E_K$ to a peak of $1.3
\times 10^{29}$~erg and the rapid decrease in free magnetic energy
of $7 \times 10^{29}$~erg.
The rate of magnetic energy decrease slows as FCS1/BCS2 transitions
from CME1's eruptive flare reconnection to the breakout reconnection
above the left arcade. The onset of FCS2 reconnection signals the
second CME's eruption and $E_K$ peak of $1.2 \times 10^{29}$~erg
during a free magnetic energy drop of $\sim 3 \times 10^{29}$~erg.
The vertical lines indicate the temporal window in the simulation
in which we will examine each of the large-scale current sheets in
detail: BCS1 red, FCS1/BCS2 green, and FCS2 blue. The black arrows
correspond to the six panels in Figure~\ref{f1}.

\section{Comparison of the Breakout and Eruptive Flare Current Sheets}
\label{s:simdiff}

The onset of reconnection in our large-scale current sheets and the
development of the tearing mode plasmoid instability can be seen
in the bottom panel of Figure~\ref{f2} where we have plotted the
number of magnetic O-type null points (magnetic islands) present
in each of our three current sheets as a function of time.
The procedure used to identify the X-type and O-type null points
is described in the Appendix of \citet{Karpen2012}. For each
simulation output file, the spatial position, type, and degree of
every magnetic null is recorded and here we have plotted only the
number of O-type nulls present in BCS1 (red), FCS1/BCS2 (green),
and FCS2 (blue).
There is an approximate correspondence between the global kinetic
energy $E_K(t)$ and the number of magnetic islands -- most visible
in the rapid rise phases of $E_K$ associated with the main CME
acceleration phase of the eruptions.

As BCS1 is stretched out and becomes unstable to tearing, the number
of magnetic islands grows from zero to $\sim$10 by 1250~s. By this
point the islands are being continually ejected into and by the
reconnection outflow exhaust and the resistive tearing of the sheet
has saturated to somewhat of a quasi ``steady-state." Continued
reconnection drives new island formation and these, in turn, are
ejected from the sheet. So from $1250 \lesssim t \lesssim 1450$~s
the rate of new island creation slightly outpaces the rate of old
island ejection and we see fluctuations around an essentially linear
trend from $\sim$5 to $\sim$12 islands present in BCS1 as it grows
(and fluctuates) in length and is pushed higher into the simulation
domain by the expanding arcade system from below.
BCS1 starts with an initial X-type null point and develops into a
current sheet via the separation of the spine lines
\citep{Syrovatskii1981,Antiochos2002,Edmondson2010} due to the
expansion of the right pseudostreamer arcade.

\begin{figure}
\center \includegraphics[width=4.25in]{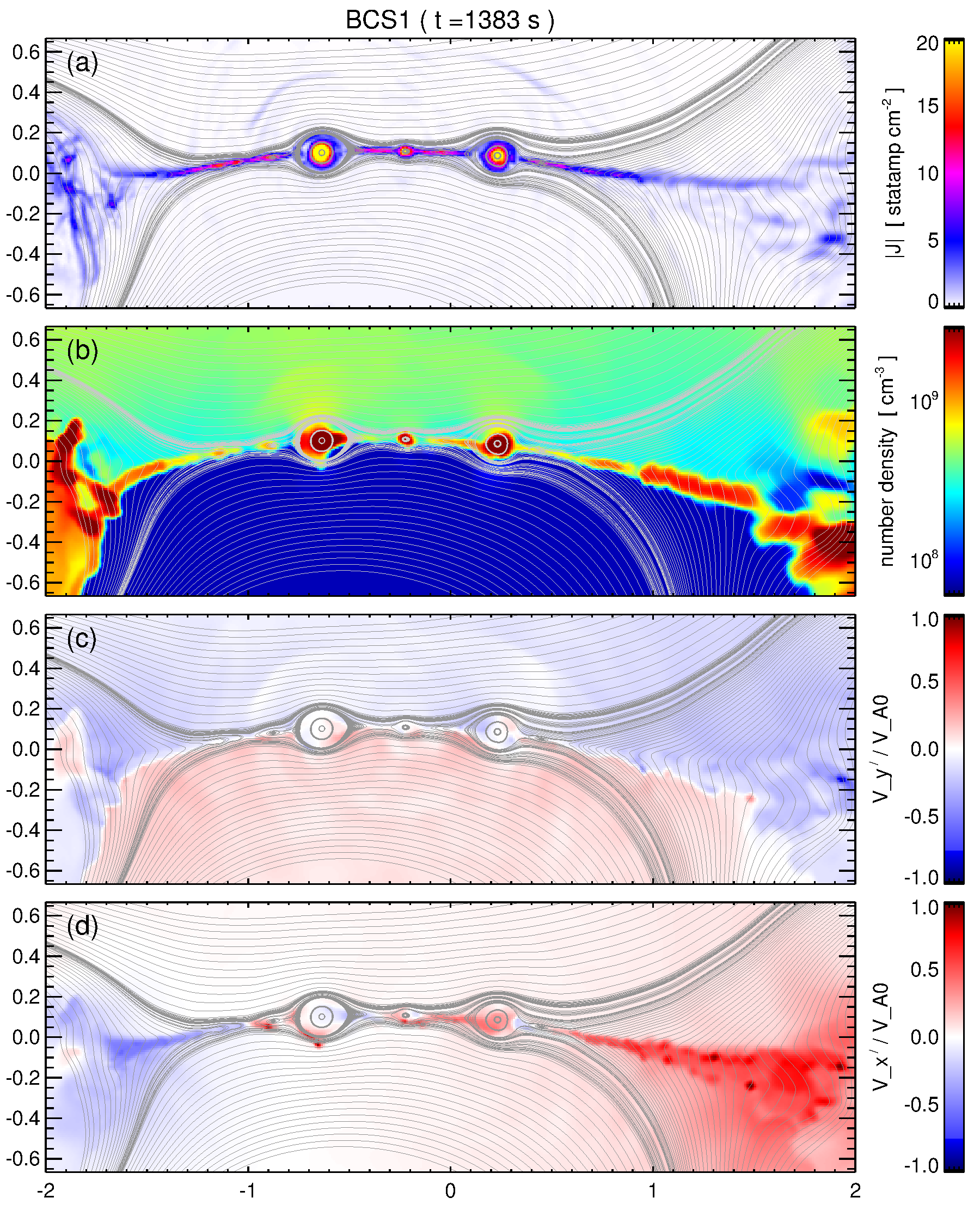}
\caption{Plasma properties of the BCS1 current sheet at $t=1383$~s:
(a) plots current density magnitude $|J|$, (b) number density
$\rho/m_p$, (c) approximate reconnection inflow $V_{y'}/V_{A0}$,
and (d) approximate reconnection outflow $V_{x'}/V_{A0}$. Representative
magnetic field lines are also plotted in each panel illustrating
the CS structure and island formation. An animation of this figure
is available as an electronic supplement to the online version ({\tt
FIGURE3\_bcs1.mp4}). \label{f3a} }
\end{figure}

Figure~\ref{f3a} shows the zoomed-in view of BCS1 and its online
animation ({\tt FIGURE3\_bcs1.mp4}) shows the temporal evolution.
Panel (a) is the electric current density $|J|$, panel (b) is the
plasma number density ($\rho/m_p)$, and panels (c) and (d) plot the
plane-of-the-sky components of the plasma velocity normalized to
the global Alfven speed, $V_{y'}/V_{A0}$ and $V_{x'}/V_{A0}$.
Here, the rectangular current-sheet centered coordinate frame ($x',
y'$) corresponds to a standard translation and rotation from the
initial simulation reference frame.
The ($x', y'$) frame locations and orientations were initially
estimated by visual inspection and then prescribed as analytic
functions of time in order to smoothly track the evolution of the
current sheets throughout the simulation domain (see
Appendix~\ref{S:Appendix1} for details).
The resulting frame orientations are such that the plasma velocity
$y'$-component in panel (c) of Figure~\ref{f3a} is approximately
aligned with the inflow into the current sheet and the $x'$-component
in panel (d) is approximately aligned with the reconnection outflow.

%
%
%
%
\begin{figure}
\center \includegraphics[width=4.25in]{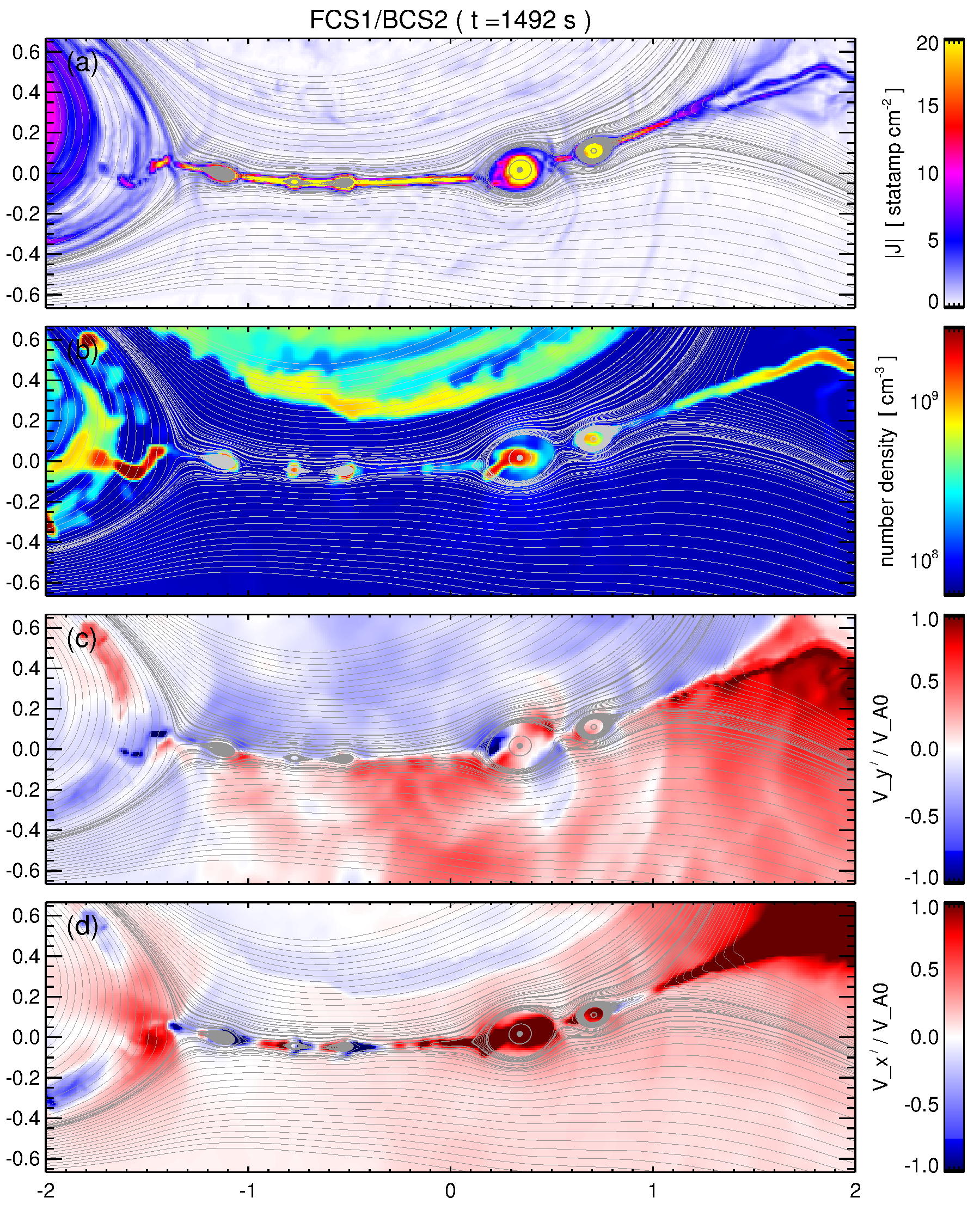}
\caption{Plasma properties and field configuration of FCS1/BCS2 at
$t=1492$~s in the same format as Figure~\ref{f3a}. An animation of
this figure is available as an electronic supplement to the online
version ({\tt FIGURE4\_fcs1.mp4}). \label{f3b} }
\end{figure}
%
\begin{figure}
\center \includegraphics[width=4.25in]{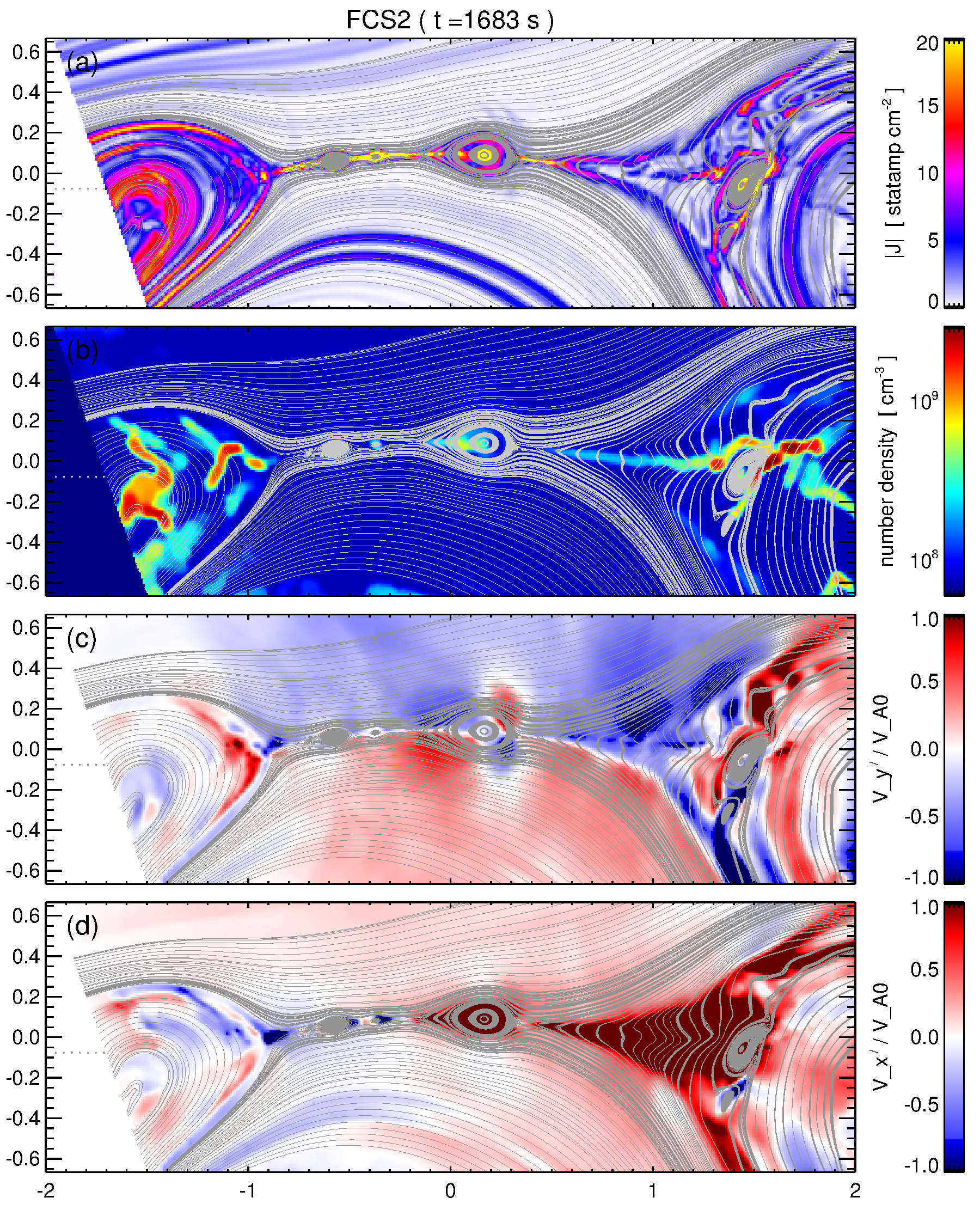}
\caption{Plasma properties and field configuration of FCS2 at
$t=1683$~s in the same format as Figure~\ref{f3a}. An animation of
this figure is available as an electronic supplement to the online
version ({\tt FIGURE5\_fcs1.mp4}). \label{f3c} }
\end{figure}
%

The evolution of FCS1/BCS2 and FCS2 start with a different onset
scenario.
For the eruptive flare current sheets, each of the pseudostreamer
arcades has a significant shear component that accumulates over the
course of the imposed boundary flows, thus developing oppositely
directed field components (in the plane of the sky) and an associated
vertical current sheet above the polarity inversion line. This is
a universal feature of all sheared arcade models
\citep{Aulanier2002,Aulanier2006,Welsch2005} that is also present
in sigmoidal field structures
\citep{Sterling2001,Canfield2007,Green2009,Savcheva2012a}, flux
emergence simulations \citep{Gibson2006,Manchester2004,Manchester2008},
and analytic and numerical treatments of flux rope models
\citep{Forbes1991a,Titov1999,Isenberg2007,Lin2009,Savcheva2012b}.

Figures~\ref{f3b}, \ref{f3c} show the zoomed-in views of FCS1/BCS2
and FCS2, respectively, in the same format as Figure~\ref{f3a}.
Their corresponding online animations, {\tt FIGURE4\_fcs1.mp4} and
{\tt FIGURE5\_fcs2.mp4}, show their respective temporal evolutions.
Again, Appendix~\ref{S:Appendix1} describes the bulk motion of the
($x', y'$) coordinate frames for each of the eruptive flare current
sheets periods.

The number of magnetic islands in FCS1/BCS2 is shown to start
moderately high ($\sim$10) and shrinks down to one by $t \sim 1460$~s
before rapidly increasing to the 15--20 range. For $t \lesssim
1460$~s the eruptive flare reconnection has not yet started in
earnest. FCS1/BCS2 exists with fluctuations in the neutral sheet
corresponding to multiple X- and O-type nulls, but there is virtually
no reconnection inflow per se. It takes the runaway arcade expansion
to disrupt the force balance sufficiently to thin the current sheet
sufficiently that numerical resistivity allows the reconnection to
proceed. Unlike the breakout reconnection, there is a significant
drop in magnetic energy so the FCS1/BCS2 outflow exceeds
the global Alfv\'{e}n speed ($|V_{x'}| \gtrsim 1.5 V_{A0}$) and
processes a significant amount of flux before the
CME eruption stretches out FCS1/BCS2 to lengths where the tearing
mode is in full effect ($t \gtrsim 1490$~s).

The second eruptive flare sheet FCS2 has the same overall dynamics
and evolution as FCS1/BCS2 for approximately the first half of its
180~s duration. Again we see the initial sheared-arcade vertical
current sheet with some islands but very little reconnection inflow
until $\sim$1670~s when the CS outflow
exceeds the global Alfv\'{e}n speed during the
impulsive phase of the eruption. FCS2 then rapidly grows beneath
the erupting flux rope, leveling off in the $\sim$10 island range.
Once $t \gtrsim 1740$~s most of the remaining free energy has been
released and the pseudostreamer arcades are able to relax towards
a state much closer to the initial potential field configuration.
The FCS2 reconnection dissipates the current density enhancement
and the CS length shrinks, i.e. the spine field lines are moving
closer together in an attempt to restore the original X-point null
topology. The reconnection becomes much smoother during this
relaxation phase and this can be seen in the number of magnetic
islands dwindling to the one or two level.

\subsection{Dimensionless Analysis: Evolution of the Global Lundquist Number}
\label{ss:lundquist}

We define the Lundquist number in the usual fashion by comparing
the annihilation timescale with the communication timescale
\textit{along} the current sheet, $S \equiv \tau_{\eta}/\tau_A$,
where $\tau_{\eta} = L^{2}/{\eta}$ and $\tau_{A} = {L}/V_{A}$. $L$
is the half-length of the current sheet, $\eta$ is the diffusion
term within the current sheet and $V_A$ the upstream Alfv\'{e}n
speed. Thus,
\begin{equation}
S \equiv \frac{\tau_{\eta}}{\tau_{A}} = \left( \frac{L^{2}}{\eta} \right) \left( \frac{V_{A}}{L} \right) = \frac{L \ V_{A}}{\eta} .
\end{equation}
In our simulation, the magnetic resistivity is purely numerical and
may be estimated by balancing the inflow speed $V_{in}$ of magnetic
flux into the current sheet against the resistive annihilation
\textit{across} the current sheet,
\begin{equation}
\tau_{in} = \tau_{\eta} \ \Longrightarrow \ \frac{\delta}{V_{in}} = \frac{\delta^{2}}{\tilde{\eta}} \ \Longrightarrow \ \tilde{\eta} = \delta \ V_{in} ,
\end{equation}
where $\delta$ is the half-thickness of the current sheet. Here we take our current sheet thickness to be the width of two computational cells at the highest grid resolution \citep[as in][]{Karpen2012}, obtaining $\delta = 10 L_0 /1024 = 0.00977L_0 = 9.77 \times 10^6$~cm.
The numerical resistivity estimates are on the order of
$10^{14}$~cm$^{2}$~s$^{-1}$ with mean values over the BCS1, BCS2/FCS1,
and FCS2 durations of $\tilde{\eta} \sim 9.5 \times
10^{13}$~cm$^{2}$~s$^{-1}$, $2.6 \times 10^{14}$~cm$^{2}$~s$^{-1}$,
and $2.9 \times 10^{14}$~cm$^{2}$~s$^{-1}$, respectively. We also
note this range is less than (i.e., compatible with) the estimate
of the strict upper-limit of the numerical resistivity, $\tilde{\eta}_{\rm
max} \equiv C \Delta x^2 / \Delta t$, where $C$ is the Courant
number, $\Delta x$ is the grid size ($\delta$) and $\Delta t$ is
the size of the computational time step (DeVore 2014, private
communication).
Substituting our numerical resistivity estimate into the Lundquist
number characterizing the current sheet yields $S = ( L V_A ) / (
\delta V_{in} )$.

We construct sheet-averaged quantities in order to estimate an
average, global Lundquist number as a function of simulation time,
\begin{equation}
\langle S(t) \rangle =  \frac{ L(t) }{\delta} \frac{ V_{A0} }{ \langle V_{in} \rangle } .
\label{e:sglob}
\end{equation}
Here, $L(t)$ is the estimate of the half-length of the sheet,
$V_{A0}$ is the global Alfven speed, and $\langle V_{in} \rangle$
represents the area average of the inflow velocities.
Appendix~\ref{S:Appendix2} provides the complete mathematical
description of CS geometries and Appendix~\ref{S:Appendix3} describes
the method used to generate the sheet-averaged estimates from the
simulation data.

%
\begin{figure}
\center \includegraphics[width=3.25in]{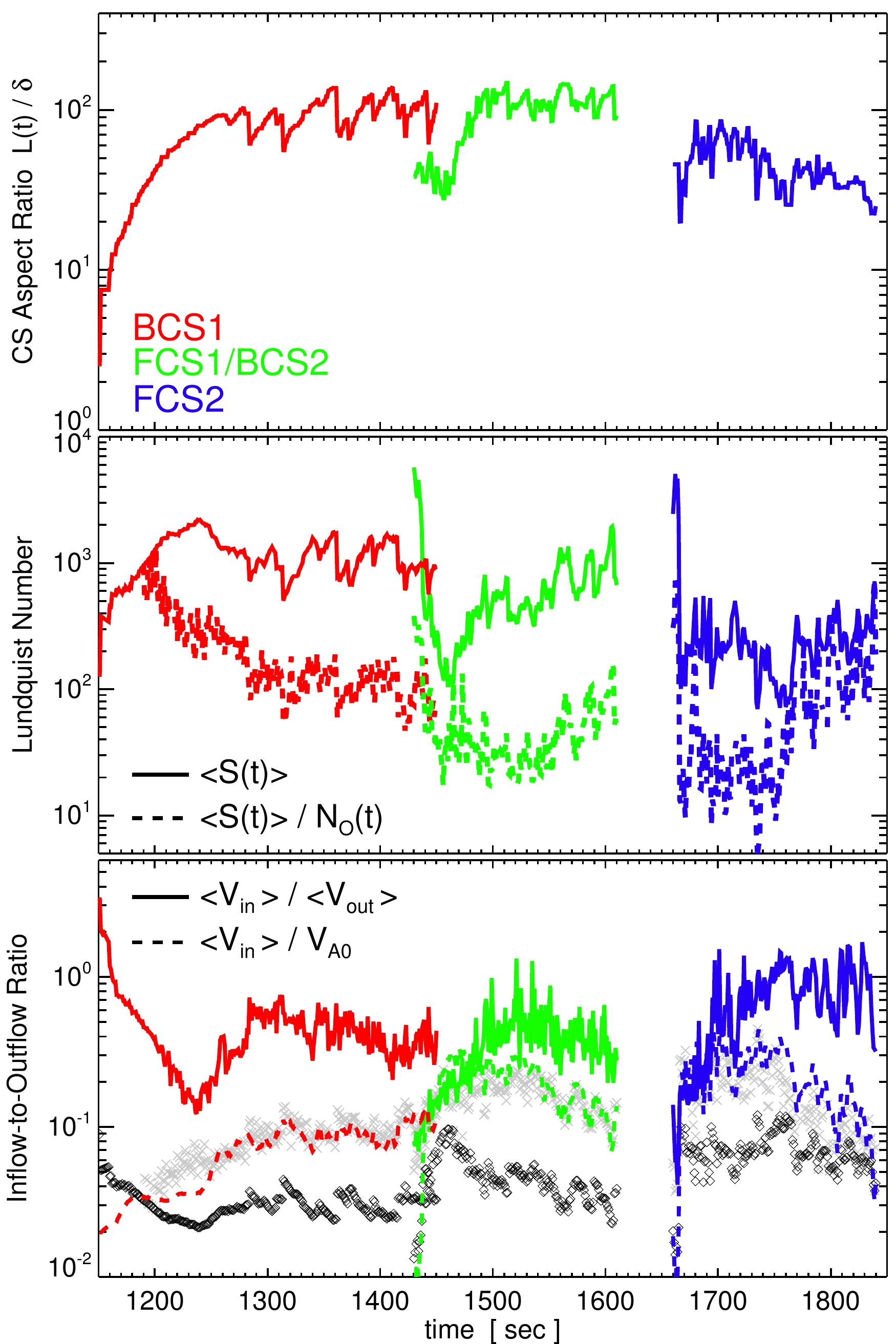}
\caption{Top panel: The length-to-width aspect ratio
$L(t)/ \delta$ for each of the three current sheets (BCS1 red;
FCS1/BCS2 green; FCS2 blue). Middle panel:  The global CS-averaged
Lundquist number ($\langle S \rangle$; solid) and the estimate of
the local CS-averaged Lundquist number ($\langle S \rangle/N_O$;
dashed). Bottom panel:  The CS-averaged inflow-to-outflow ratio
($\langle V_{in} \rangle / \langle V_{out} \rangle$; solid), the
inflow-to-global Alfv\'{e}n speed ratio ($\langle V_{in} \rangle /
V_{A0}$; dashed), the theoretical Sweet-Parker scaling ($S^{-1/2}$;
black diamonds), and the plasmoid-modified Sweet-Parker scaling
($S^{-1/2}N_O^{1/2}$; gray crosses). See text for details. \label{f5a}
}
\end{figure}
%


The top panel of Figure~\ref{f5a} plots the time evolution of the
average CS aspect ratio $ L(t) / \delta$ for BCS1 (red), FCS1/BCS2
(green), and FCS2 (blue). The BCS1 aspect ratio clearly shows the
rapid formation and elongation of the CS as the spine lines separate
and a gradual leveling-off at $\sim$80:1 by $t \gtrsim 1250$~s.
From this point onward the BCS1 aspect ratio fluctuates around this
level, ranging from 60:1--100:1. The sharp decreases in length
correspond to ejections of large plasmoids and typically the sheet
lengthens again until the next large plasmoid ejection. The FCS1/BCS2
aspect ratio shows essentially the same large-scale evolution.
Starting with the onset of flare reconnection ($t\sim1460$~s), the
FCS1/BCS2 aspect ratio grows from 30:1 to $\sim$100:1 before
leveling-off. FCS2 on the other hand, while showing similar rapid
growth starting from its flare reconnection onset ($t\sim1670$~s),
starts from a ratio of $\sim$20:1 but levels-off earlier at $\sim$60:1
before gradually shrinking back towards the $\sim$20:1 level over
the course of $1750 \lesssim t \lesssim 1840$~s.

The middle panel of Figure~\ref{f5a} plots the \emph{global}
CS-averaged Lundquist number $\langle S(t) \rangle$, given by
Equation~\ref{e:sglob}, as a function of time in each of the three
sheets (BCS1 red; FCS1/BCS2 green; FCS2 blue).
The behavior of $\langle S(t) \rangle$ shows a similar evolution
to the CS aspect ratio. The BCS1 Lundquist number increases rapidly
to $10^3$ through 1200~s during the elongation of the CS and then
continues to increase more gradually until  it reaches $2 \times
10^3$ by 1250~s. The Lundquist number then fluctuates around
approximately $10^3$ through the rest of the CS evolution.
The FCS1/BCS2 current sheet has a comparable average Lundquist
number of $\sim$800--1000 but transitions from higher values due
to an initially near-zero inflow velocity. Once plasmoid-unstable
reconnection has started, FCS1/BCS2 also fluctuates around this
average. The average FCS2 $\langle S(t) \rangle$ is a bit lower,
at $\sim$200--400 as the current sheet dissipates.

The middle panel of Figure~\ref{f5a} also plots $\langle S(t) \rangle
/ N_O(t)$ as dotted lines corresponding the global Lundquist number
divided by the number of O-type magnetic null points, $N_O$, from
Figure~\ref{f2}. This is an estimate of the CS-averaged \emph{local}
Lundquist number for the CS intervals between islands. \citet{Cassak2009b},
\citet{Daughton2009}, \citet{Uzdensky2010}, and others have argued
that during the non-linear phase of the instability these secondary
sheets also reach the tearing threshold, go unstable, and start to
generate islands. Our static computational grid imposes the minimum
CS thickness of $\delta$ so we do not resolve the thinning of the
secondary sheets in this simulation. Thus, in a statistical sense,
the ``steady state'' number of magnetic islands, $N_O \sim 10$, and
global $S \sim 10^3$ imply secondary sheets with $S_{local} \sim
10^2$.

The canonical critical Lundquist number is typically taken as
$\sim$10$^4$ \citep[e.g.,][and references
therein]{Biskamp1986,Samtaney2009,Huang2010,Loureiro2012,Murphy2013} but
again, the plasmoid instability has been shown to develop over a
wide range of values depending on the specific simulation details.
Under steady-state driving (inflow) conditions, the \citet{Edmondson2010}
MHD simulations were plasmoid unstable and comfortably in the
nonlinear regime at $S \sim 1.2\times10^3$, consistent with the
\citet{Ni2010} and \citet{Shen2011} results.
Here, our \emph{global} Lundquist number is similar to
\citet{Edmondson2010} but our inflow-to-outflow ratio is higher and
our magnetic island plasmoids appear to cover a much greater dynamic
range in size (discussed further in Section~\ref{ss:distributions}).

\begin{figure}
\center \includegraphics[width=3.25in]{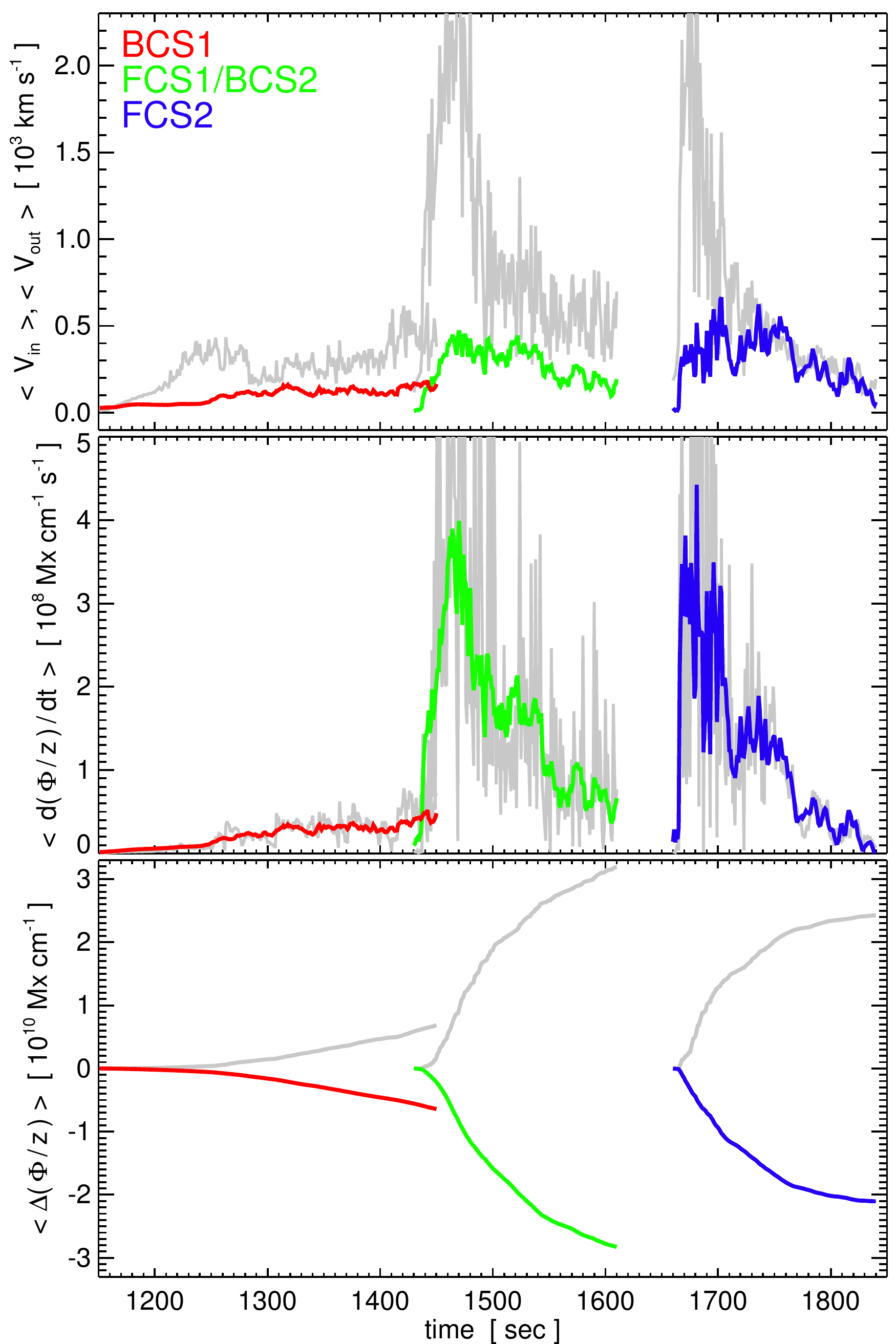}
\caption{Top panel: Current sheet-averaged inflow velocities $\langle
V_{in} \rangle$ (BCS1 red; FCS1/BCS2 green; FCS2 blue) and outflow
velocities $\langle V_{out} \rangle$ (gray). Middle panel: Estimate
of the CS-averaged reconnection rate magnitude $\langle d(\Phi/z)/dt
\rangle$ calculated from the $z$-component of $(\mathbf{V} \times
\mathbf{B})$, with the flux into the sheet plotted as the colored
lines and flux out of the sheet plotted as gray. Bottom panel: Total
change in flux content $\langle \Delta(\Phi/z) \rangle$ due to the
signed reconnection rates above. \label{f5b} }
\end{figure}

The bottom panel of Figure~\ref{f5a} shows two different measures of the 
CS-averaged inflow-to-outflow ratio. The solid line shows 
$\langle V_{in} \rangle / \langle V_{out} \rangle$ in the standard color scheme. 
This ratio is calculated directly from the $\langle V_{in} \rangle$ and 
$\langle V_{out} \rangle$ profiles from the top panel of Figure~\ref{f5b}.
Here we also plot the inflow-to-global Alfv\'{e}n speed ratio 
$\langle V_{in} \rangle / V_{A0}$ as the dashed colored lines. 
It is common in reconnection theory to simply equate the outflow 
speed to $V_{A0}$. However, it is also a common feature of MHD 
simulations to have reconnection outflow be a fraction (albeit a 
substantial fraction, i.e. $\sim$50\%) of the global Alfv\'{e}n speed 
\citep[e.g., as in][]{Karpen1995,Murphy2010,Edmondson2010,Shen2011}.
The theoretical Sweet-Parker inflow-to-outflow scaling, calculated
from the simulation's global Lundquist number as $\langle S(t)
\rangle^{-1/2}$, is plotted as black diamonds and the
\citet{Cassak2009b} plasmoid-modified Sweet-Parker scaling,
$S_{local}^{-1/2} = \langle S(t) \rangle^{-1/2}N_O^{1/2}$, is plotted
as gray crosses.

For BCS1, the $\langle V_{in} \rangle / \langle V_{out} \rangle$ ratio 
decreases as the CS elongates until significant outflow develops by $t \gtrsim 1200$~s. From $1250
\lesssim t \lesssim  1300$~s there is a transition from the 0.1--0.2
range to $\sim$0.4 where it remains for most of its duration.
The FCS1/BCS2 inflow-to-outflow behavior is very similar. Once the
eruptive flare reconnection starts, the inflow-to-outflow ratio
transitions from the same $\sim$0.1 range to the $\sim$0.4 range
where it remains for its duration. After the first CME eruption,
FCS1/BCS2 is acting as the second large-scale breakout sheet so the
agreement with the BCS1 results could be expected, but that the
initial flare reconnection phase of FCS1/BCS2 so closely resembles
the initial phase of BCS1 highlights the role of the onset and
development of the plasmoid instability in increasing the overall
inflow-to-outflow ratio ``reconnection rate."
The FCS2 ratio also shows this transition in the first half of its
evolution, but as discussed earlier, the second half of the FCS2
evolution is a different physical situation than the other two CSs.
The inflow-to-outflow ratio remains ``high'' ($\sim$0.4) but the
sheet itself is shrinking and both the inflow and outflow speeds
are decreasing significantly.

There are two important features of the Figure~\ref{f5a} inflow-to-outflow results. 
First, throughout the entire simulation and for every CS, our 
$\langle V_{in} \rangle / \langle V_{out} \rangle$ ratio remains 
significantly higher than the classical Sweet-Parker $\langle S \rangle^{-1/2}$ 
scaling. Second, our $\langle V_{in} \rangle / V_{A0}$ ratio matches the 
\citet{Cassak2009b} plasmoid-modified Sweet-Parker scaling $S_{local}^{-1/2}$ 
almost exactly. The differences between the 
$\langle V_{in} \rangle / \langle V_{out} \rangle$ and 
$\langle V_{in} \rangle / V_{A0}$ measures arise from the bulk 
plasma velocities adjusting in tandem to keep the 
inflow-to-outflow ratio large enough to accomplish 
the necessary flux transfer imposed by the global evolution of the 
sympathetic CME eruptions. The magnitudes of both the inflow 
and outflow velocities vary significantly depending on what role 
the reconnection is playing in the global eruption scenario: 
$\langle V_{out} \rangle$ exceeds 2000~km~s$^{-1}$ during the 
eruptive flare but is only $\sim$500~km~s$^{-1}$ during the 
breakout reconnection in BCS1 and the later breakout phase 
of FCS1/BCS2. By late in the FCS2 evolution the 
$\langle V_{in} \rangle / \langle V_{out} \rangle$ ratio approaches 
unity but this only corresponds to 100$-$200~km~s$^{-1}$ speeds 
and comparatively low flux transfer rates.

%

%
%

\subsection{Evolution of Current Sheet Reconnection in Physical Units}
\label{ss:rxnrate}

The top panel of Figure~\ref{f5b} plots CS-averaged $\langle V_{in}
\rangle$ (colored lines) and $\langle V_{out} \rangle$ (gray lines)
in units of $10^3$ km~s$^{-1}$. The ``steady state" inflow speed
in BCS1 $\sim$100~km~s$^{-1}$ while the outflow speeds are
$\sim$250~km~s$^{-1}$ with intermittent periods up to
$\sim$400~km~s$^{-1}$. The FCS1/BCS2 and FCS2 inflow and outflow
speeds have a different temporal character than the BCS1 evolution.
FCS1/BCS2 and FCS2 inflow speeds are $\sim$300--400~km~s$^{-1}$ for
80--100 seconds before dropping back down to 100--200~km~s$^{-1}$.
The CS-averaged flare reconnection outflows rapidly reach
$\gtrsim$2000~km~s$^{-1}$ during the period of maximum inflow speed,
and drop down to $\sim$600~km~s$^{-1}$ for FCS1/BCS2 and
$\sim$200~km~s$^{-1}$ for FCS2 as the last of the strong currents
dissipate.

The middle panel of Figure~\ref{f5b} plots the time rate of change
of the current-sheet averaged flux processed by reconnection, i.e.,
the $z$-component of the plane-of-the-sky $\mathbf{V} \times
\mathbf{B}$ (see Appendix~\ref{S:Appendix3}). The colored lines
indicate $\langle d(\Phi/z)/dt \rangle$ calculated with the inflow
area averaging procedure and the gray lines correspond to $\langle
d(\Phi/z)/dt \rangle$ calculated with the outflow area averaging
procedure.
The lower panel of Figure~\ref{f5b} plots the time integrated
reconnection rates from the above panel, $\langle \Delta(\Phi/z)
\rangle = \int dt \langle d(\Phi/z)/dt \rangle$, to show the
cumulative flux processed through each of the CS reconnection
regions. Again, the total change in flux due to the inflow averaging
is shown as the colored lines and the outflow averaging is shown
in gray.

\citetalias{Lynch2013} discussed how, despite the topological
similarity between the breakout and flare current sheets, the flare
reconnection generates much greater flux transfer rates and
inflow/outflow velocity magnitudes because of its fundamentally
different roles in the global eruption scenario. The $\langle
d(\Phi/z)/dt \rangle$ and $\langle \Delta(\Phi/z) \rangle$ results
presented here -- calculated directly from the CS regions of the
simulation -- can be compared to \citetalias{Lynch2013} Figures~4(c)
and 4(b), respectively, where we calculated the same quantities
from the evolution of the global flux content in each of the
pseudostreamer arcades and newly-formed flare arcades.
The peak reconnection rates in FCS1/BCS2 and FCS2 ($\sim$$3 \times
10^{8}$ Mx~cm$^{-1}$~s$^{-1}$) are comparable to the \citetalias{Lynch2013}
version, and in both simulations, the maximum flare reconnection
rates correspond to approximately 10 times the BCS1 reconnection
rate.
Likewise, the total flux transfered -- measured in \citetalias{Lynch2013}
by following the pseudostreamer arcade separatrices in time and
integrating the normal field at the simulation's lower boundary --
agrees reasonably well: $\sim$$6 \times 10^{9}$~Mx~cm$^{-1}$ (BCS1),
$\sim$$3 \times 10^{10}$~Mx~cm$^{-1}$ (FCS1/BCS2), and $\sim$$2
\times 10^{10}$~Mx~cm$^{-1}$ (FCS2).

It is worth emphasizing that the method by which the reconnection
rate was derived in \citetalias{Lynch2013} is essentially the
procedure used to estimate the reconnection rate from solar
observations: one measures the area swept up by the flare ribbons
in time and calculates the amount of magnetic flux from the underlying
photospheric flux distribution \citep[e.g.,][]{Forbes2000, Qiu2002,
Qiu2004, Jing2005, Kazachenko2012}.
The method presented here for deriving the reconnection rate looks
exclusively at the field and plasma evolution at the coronal CS
regions that are currently extremely difficult to observe.
The agreement here with the \citetalias{Lynch2013} reconnection
fluxes is a trivial simulation result (magnetic flux is conserved
even when the CS is highly structured, dynamic, and plasmoid unstable)
but the implications for the observational flare-ribbon technique
as a robust measure of the flux participating in the eruptive flare
reconnection is encouraging.

\section{Small-Scale Structure in Plasmoid Unstable Current Sheets}
\label{s:islands}

\subsection{Distributions of Magnetic Island Area, Mass, and Flux Content}
\label{ss:distributions}

\begin{figure}
\center \includegraphics[width=3.25in]{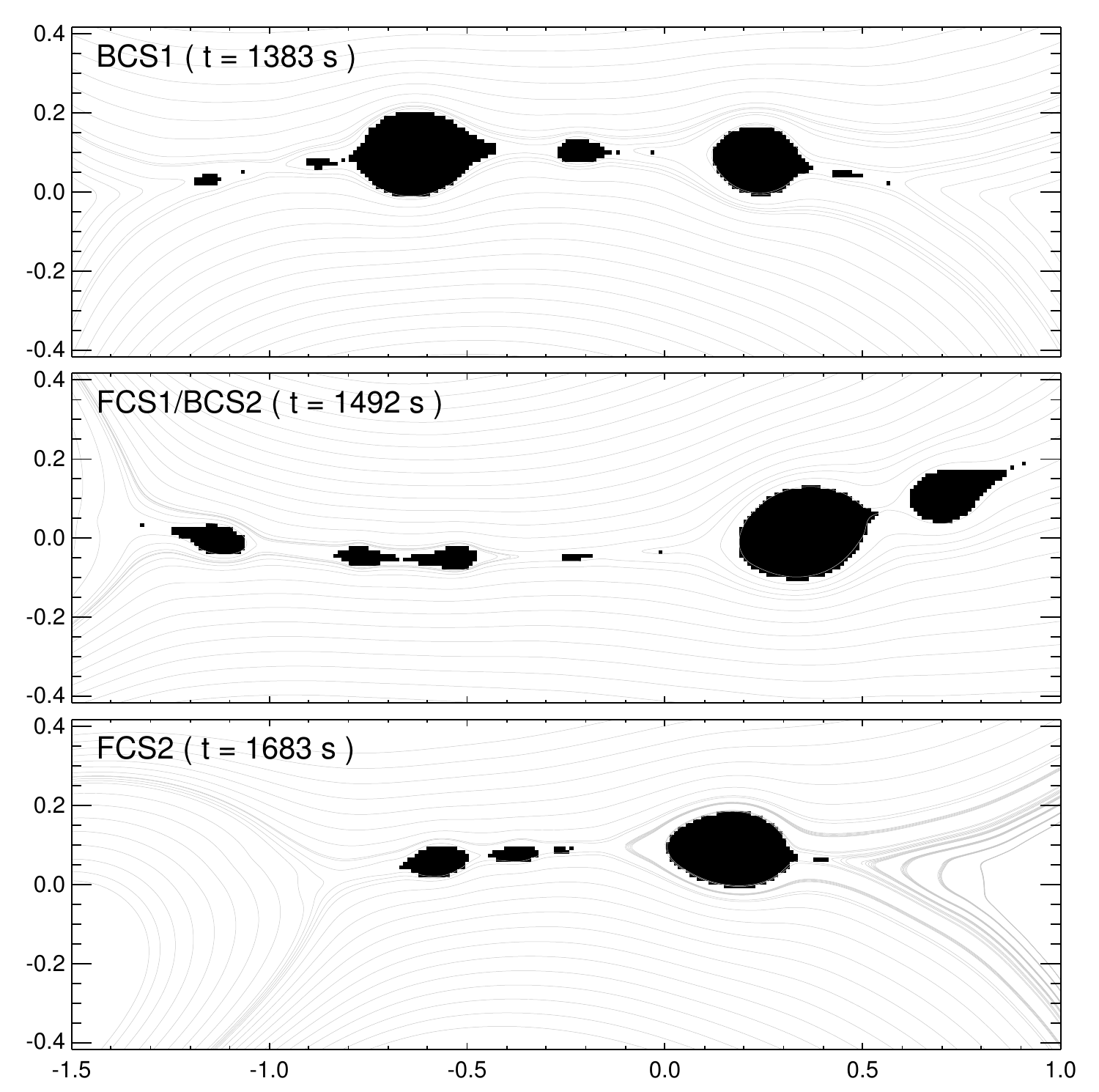}
\caption{Pixel masks for determining magnetic island area, mass,
and flux content for each of the simulation times shown in
Figures~\ref{f3a}--\ref{f3c}. \label{f11.5} }
\end{figure}

To calculate the area of our plasmoid magnetic islands and their
mass and flux content, we create a pixel mask associated with each
O-type null in every $(x', y')$ simulation output frame. These masks
are constructed by integrating a set of magnetic field lines between
the adjacent X-points and only plotting those that belong to the
topological domain of the magnetic island, i.e. that do not exceed
the spatial position of the bounding X-points. Figure~\ref{f11.5}
shows a representative pixel mask, one for each CS at the simulation
times of Figures~\ref{f3a}--\ref{f3c}. The island area $A$ is
calculated simply by summing the number of non-zero pixels in the
mask and multiplying by the pixel area $\delta^2 = 9.54 \times
10^{13}$~cm$^{2}$.
The island pixel masks multiplied by the mass density and magnetic
field components allow us to construct the plasma and flux contents
associated with each island (index $o$):
\begin{eqnarray}
A^{o} = \int dA^o &=& \sum_{ij} \delta^2,\\
(m/z)^{o} = \int \rho \; dA^o &=& \sum_{ij} \delta^2\;  \rho_{ij} ,\\
(\psi/z)^{o} = \int \left( \mathbf{B} \cdot \mathbf{\hat{e}_2} \right) \; dx' &\simeq& \onehalf \sum_{i} \delta \; | \left( \mathbf{B} \cdot \mathbf{\hat{e}_2} \right) |_{i, \; j={\rm fixed}}  \label{e13} \\
= \int \left( \mathbf{B} \cdot \mathbf{\hat{e}_1} \right) \; dy' &\simeq& \onehalf \sum_{j} \delta \; | \left( \mathbf{B} \cdot \mathbf{\hat{e}_1} \right) |_{i={\rm fixed}, \; j}  \label{e14}.
\end{eqnarray}
\\
Both the plasmoid mass ($m/z$) and plane-of-the-sky flux ($\psi/z$)
represent per unit length values, gm~cm$^{-1}$ and Mx~cm$^{-1}$,
respectively.
Here, $i$, $j$, denote the pixel indices in the ($x'$, $y'$) frame.
The island flux content is estimated as the mean of the values
obtained from the line integral of the $\mathbf{B} \cdot
\mathbf{\hat{e}_2}$ magnitude along $dx'$ through the island and
from the line integral of the $\mathbf{B} \cdot \mathbf{\hat{e}_1}$
magnitude along $dy'$ through the island. For under-resolved islands
($A \sim$~a few pixels), the planar flux estimates are necessarily
approximate. However, for well-resolved islands ($A \gtrsim
10^2$~pixels), Equations~\ref{e13} and \ref{e14} yield similar
numerical values as expected.

%

\begin{figure*}
\center \includegraphics[width=6.5in]{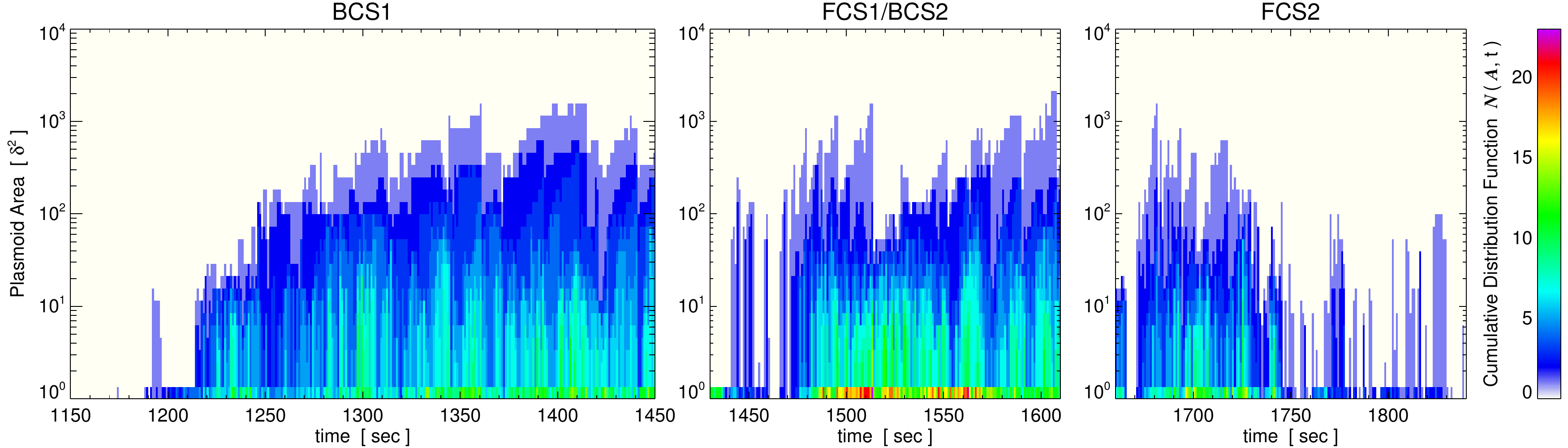}
\caption{Temporal evolution of the cumulative distribution function
of plasmoid area $N(A,t)$ for each of the three current sheets.
$N(A,t)$ represents the number of total plasmoids of area $A$ or
greater (in units of the pixel area $\delta^2 = 9.54 \times
10^{13}$~cm$^{2}$). \label{f12} }
\end{figure*}

The probability distributions functions for plasmoid width, flux,
and mass content are all derived from their respective cumulative
distribution functions \citep[][]{Uzdensky2010,Shen2013}.
For the magnetic island plasmoid area, the cumulative distribution
function $N(A,t)$ measures the number of plasmoids of area $A$ or
greater at time $t$.
Figure~\ref{f12} plots the cumulative area distribution as a function
of time for each CS where we have used 30 bins spaced uniformly in
$\log{A}$ over the range $A \in [1,10^4]$ in units of pixel area
$\delta^2$. The cumulative distribution functions for mass and flux
content are likewise calculated and binned over the ranges $m/z \in
[10^{-2},10^{6}]$~gm~cm$^{-1}$ and $\psi/z \in
[10^{5},10^{11}]$~Mx~cm$^{-1}$.

%
%

%
%
\begin{figure*}[tbh]
\center \includegraphics[width=6.5in]{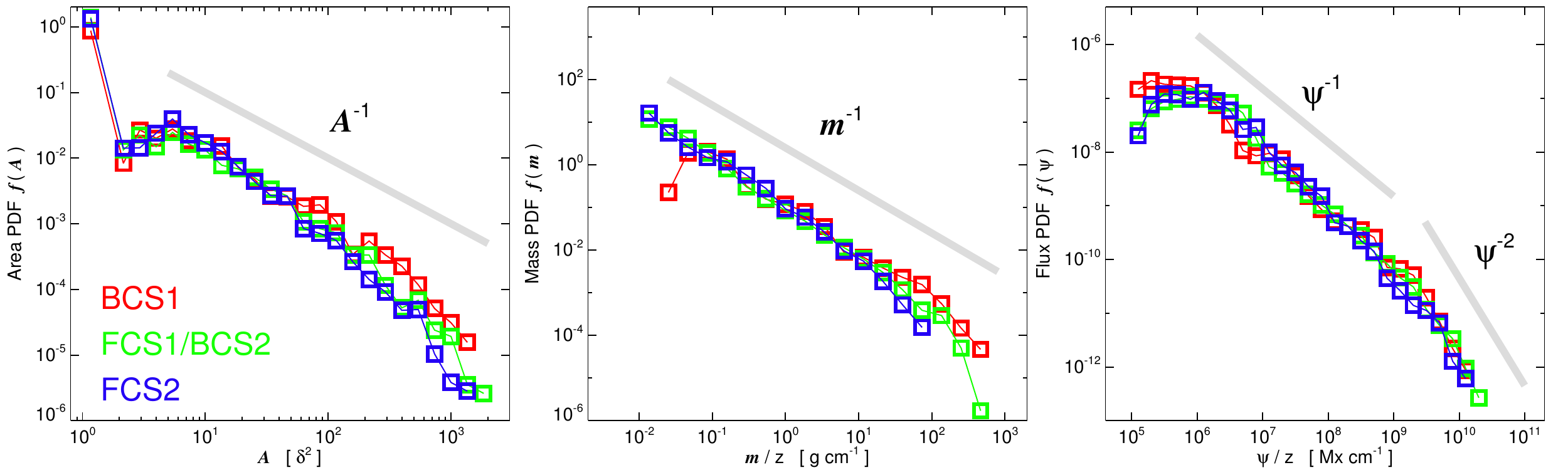}
\caption{The normalized probability distribution function (PDF) of
plasmoid area $f(A)$ (left), mass content $f(m)$ (middle), and
average planar flux $f(\psi)$ (right) for each of the three current
sheets BCS1 (red), FCS1/BCS2 (green), and FCS2 (blue). Each PDF is
constructed from their respective cumulative distribution functions
summed over each CS time period. Power law slopes are shown for
reference in light gray.  \label{f13} }
\end{figure*}

We sum over each of the CS durations ($\tau$) to obtain the cumulative,
time-integrated total, $N_\tau(A)$.
The probability distribution functions are then obtained as $f(A)
= -dN_\tau(A)/dA$ and correspond to the number of plasmoids per
unit area for an area $A$. Here we apply the additional normalization
$\Sigma f(A)\Delta A = 1$ to ease the comparison between our three
cases that generate different numbers of total plasmoids.
The left panel of Figure~\ref{f13} shows the normalized probability
density function $f(A)$ calculated for each CS. The middle panel
plots the mass density PDF $f(m)=-dN_\tau(m/z)/d(m/z)$.
The right panel of Figure~\ref{f13} shows the flux distribution
calculated from $f(\psi)=-dN_\tau(\psi/z)/d(\psi/z)$. We also plot
reference power law slopes for each of the Figure~\ref{f13} panels
as thick light gray lines.

In general, the mass and flux distributions reflect the plasmoid
area distribution in each of the three current sheets. This is to
be expected if the mass density and upstream field strengths are
essentially uniform over the current sheet. For example, if $f(A)
\sim A^{-1}$ and the mass content is $m \sim \rho_0 A$, it follows
that $f(m) \sim m^{-1}$.  The island area is proportional to an
(approximate) island width of $w^2$ and therefore our $f(A) \lesssim
A^{-1}$ in Figure~\ref{f13} implies an equivalent $f(w) \lesssim
w^{-2}$ scaling. Estimating a planar flux content of $\psi \sim B_0
w$ yields a flux distribution of $f(\psi) \sim \psi^{-2}$.

\citet{Uzdensky2010} argue that in a stochastic, self-similar
plasmoid chain, the fluxes should scale as $f(\psi) \sim \psi^{-2}$
and the island widths, $f(w) \sim w^{-2}$. However, \citet{Huang2012}
showed numerical simulations that produce a $\psi^{-1}$ scaling for
the flux distribution in magnetic islands, and \citet{Fermo2010}
predict an exponential distribution function. \citet{Shen2013}
examined the flux and width distributions of magnetic islands and
found qualitatively similar results to \citet{Loureiro2012}: the
width and flux distributions had slopes between ${-1}$ and ${-2}$
that steepened towards $-2$ for the larger values.
Observationally, the distribution of plasmoids in LASCO C2 coronagraph
data (density enhancement ``upflows'' in a post-CME radial plasma
sheet) has been investigated by \citet{Guo2013} who found a log-normal
shape that was also consistent with an exponential decay for plasmoid
widths $\gtrsim 50$~Mm. Lower in the corona, the collimated voids
(density depletions) seen by \citet{McKenzie2011} in flare arcade
plasma sheets -- called Supra-Arcade Downflows (SADs)  -- also have
an apparent log-normal distribution and recent simulations by
\citet{Cassak2013b} were used to examine the relationship between
SADs and flare reconnection outflows. Likewise, \citet{Fermo2011}
showed that Flux Transfer Events observed in the magnetotail by
\emph{Cluster} were also consistent with a log-normal and/or an
exponential decay for widths $\gtrsim 4$~Mm.

Our results are consistent with the flatter portion of the
\citet{Loureiro2012} and \citet{Shen2013} distributions and the
$\psi^{-1}$ scaling found by \citet{Huang2012}. It may be that our
choice to dynamically ``shorten" the current sheet boundaries over
the plasmoids as they approach the end of the sheet means we
under-sample the largest plasmoid values and enhance the steepening
of the distribution slope at the highest values.
Despite this possible selection effect, it is interesting that the
largest plasmoids we do count regularly ($A \gtrsim 10^2$~pixels)
occur in each of the three current sheets. \citet{Uzdensky2010} and
\citet{Loureiro2012} have discussed ``monster plasmoids" that grow
out of the combination of continued reconnected flux accumulation
and the coalescence of smaller plasmoids. In our results, the largest
plasmoids also reach ``macroscopic" sizes, i.e., of order of 10--20\%
of the total current sheet length ($L \sim 200\delta$), and from
Figure~\ref{f12}, are seen to appear, get ejected, and re-appear
regularly during a large fraction of both the BCS1 and FCS1/BCS2
time intervals. FCS2 also generates a couple of large plasmoids
during the impulsive phase of its eruptive flare before the system
relaxation smooths out the reconnection.

\subsection{Evolution of the Guide Field Component at the Current Sheets}
\label{ss:gf}

%
\begin{figure*}[t]
\center \includegraphics[width=6.5in]{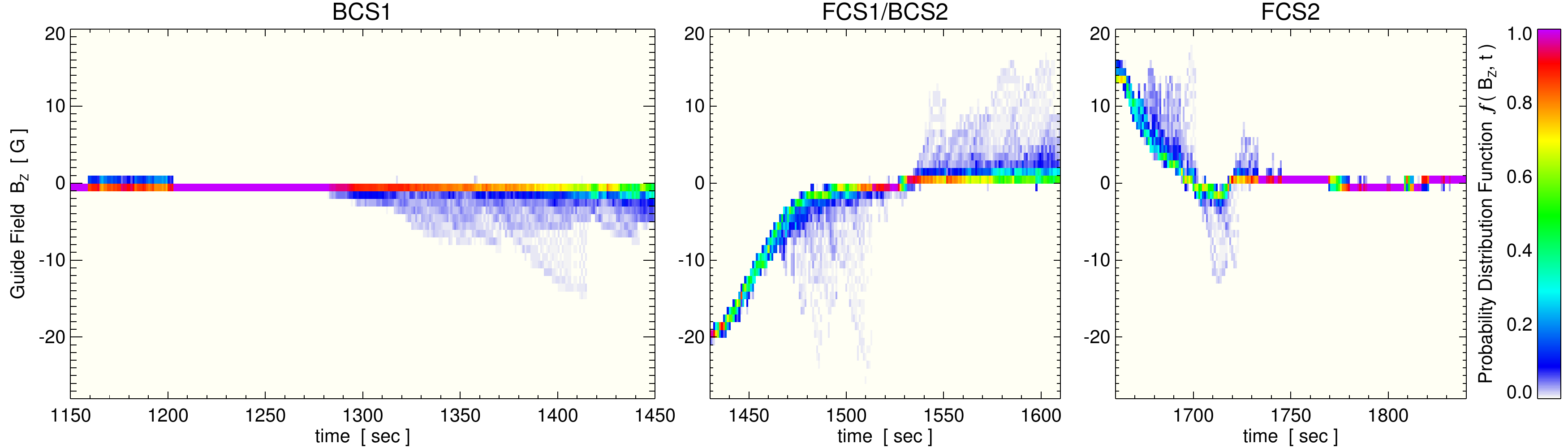}
\caption{Temporal evolution of the distribution of the guide-field
component $B_z$ along each current sheet. \label{f15} }
\end{figure*}

Given the differences in the location and the role each current
sheet plays in the sympathetic eruption scenario, it is not necessarily
obvious that the reconnection and plasmoid properties would agree
as much as they do. We calculated the distribution of guide-field
flux $(\psi_z)^o = \int B_z dA^o$ in each island as in
$\S$\ref{ss:distributions} and found that $f(\psi_z) \sim \psi_z^{-1}$
for exactly the same reasoning ($\psi_z \propto A$ for constant
$B_z$). However, $B_z$ does undergo large-scale changes at the CS
during reconnection because of the structure and evolution of the
sheared fields associated with the eruptions.

Figure~\ref{f15} plots the distribution of the guide-field $B_z$
sampled along each of the CS arcs. Here we have constructed $f(B_z,t)$
by sampling and binning the $B_z(x',y')$ values along the
Appendix~\ref{S:Appendix2} CS arc fits at each output time.
The BCS1 $f(B_z,t)$ distribution is strongly peaked at $B_z \sim
0$ but for $t \gtrsim 1300$~s, the reconnecting flux starts to
include some of the expanding right-side arcade's shear component.
The process of magnetic island formation also concentrates a
relatively weak guide-field component into localized, relatively
strong peaks, as seen in the tail of the $f(B_z,t)$ extending through
$-10$~G.
FCS1/BCS2 starts deep in the shear channel with $f(B_z,t)$ highly
peaked at $-20$~G and shows a smooth evolution toward zero as the
sheared flux reconnects during the eruptive flare and CME formation.
Again, once the guide-field magnitude drops below 10~G the plasmoid
formation process broadens the tail of the distribution to larger
$B_z$ values. The FCS1/BCS2 transition between flare reconnection
for the first CME and breakout reconnection for the second CME is
obvious -- the guide field component switches sign from negative
to positive $B_z$ over $1510 \lesssim t \lesssim 1550$~s indicating
the sheared flux of the expanding pseudostreamer left-side arcade
is now being processed through the CS.
After $1550$~s, the FCS1/BCS2 distribution looks similar to BCS1
but with the opposite sign.
The FCS2 guide-field distribution starts strongly peaked at the
shear channel value of $\sim$15~G and develops the same characteristic
distribution broadening as the peak moves towards zero. There is
some oscillation in the sign of the weak guide-field values (and
thus the extended distribution tails) before FCS2 settles down and
the magnetic free energy has been expended.

\subsection{Spectral Properties of Current Sheet Magnetic Fluctuations}
\label{ss:wavelets}

\subsubsection{Wavelet Analysis of Plasmoid Structures}

We have performed a wavelet spectral analysis to characterize the
spatial scales and power spectra of the magnetic field fluctuations
associated with the reconnection-generated magnetic islands in each
of our three current sheets. Wavelet analyses have an advantage
over traditional spectral methods (Fourier transform) by being able
to isolate both large timescale and small timescale periodic behavior
that occur over only a subset of the time series \citep[see][and
references therein]{Edmondson2013b}.
We sample the simulation data along the CS arc obtained via the
method of Appendix~\ref{S:Appendix2} using $x'$ as our spatial
position parameter to obtain a quantity $Q(x') = (B^2(x')/(8\pi))^{1/2}$.
The wavelet transform of $Q$ is defined as
\begin{equation}
W_Q(x',X) = \int Q(\xi)\Psi^*(\xi,x',X)d\xi
\end{equation}
with the Morlet family waveform given by
\begin{equation}
\Psi^*(\xi,x',X) = 
\frac{\pi^{1/4}}{|X|^{1/2}} \exp\left[ i \omega_0 \frac{(\xi-x')}{X} \right] \exp\left[ - \frac{(\xi-x')^2}{2X^2} \right].
\end{equation}

\noindent In our application, $\xi$ is the integration variable,
$x'$ is the position in the CS-centered rectangular frame, and $X$
is the wavelet spatial scale (corresponding to inverse spatial
frequency). The non-dimensional frequency parameter $\omega_0 = 6$
corresponds to approximately three oscillations within the Gaussian
envelope.
The rectified wavelet power spectra is obtained by the square of
the wavelet transform amplitude
\begin{equation}
\mathcal{P}_Q(x',X) = |X|^{-1}|W_Q(x',X)|^2
\end{equation}
where we have employed the \citet{Liu2007} frequency scaling to
correct for the inherent low-frequency (large spatial scale) bias
due to the width of the wavelet filter in frequency space.

%
\begin{figure*}[!tbh]
\includegraphics[width=6.50in,height=1.5in]{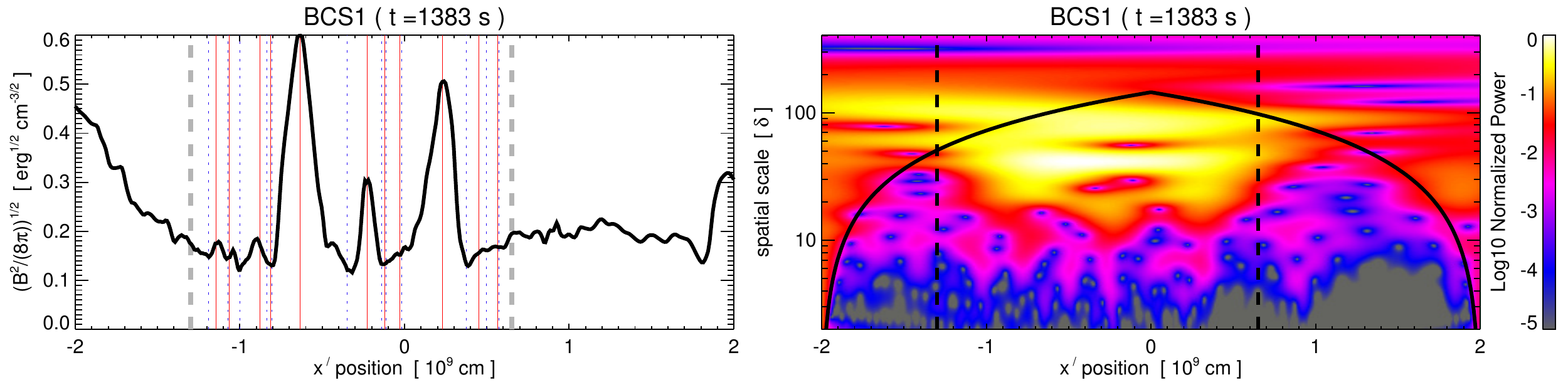}
\includegraphics[width=6.50in,height=1.5in]{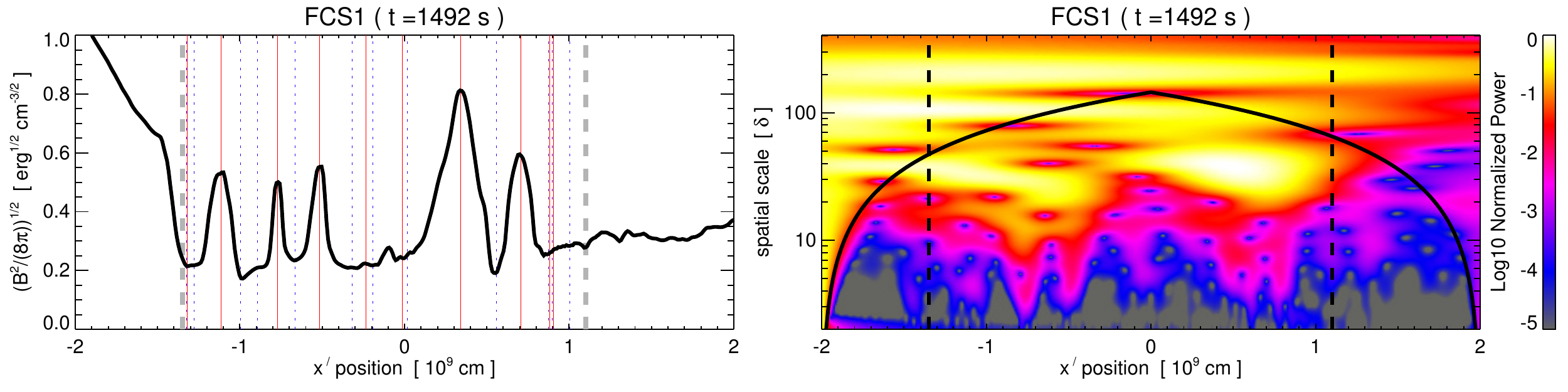}
\includegraphics[width=6.50in,height=1.5in]{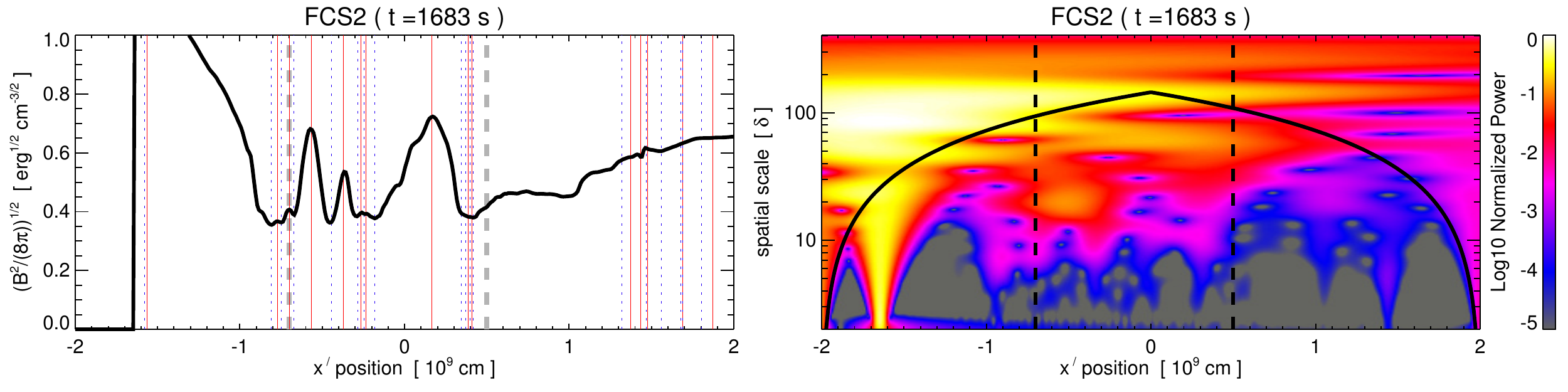}
\caption{Left column: Sampling of $(B^2/(8\pi))^{1/2}$ along the
curvilinear arc fits for BCS1 (top row), FCS1/BCS2 (middle row),
and FCS2 (bottom row). The CS boundaries are shown as vertical
dashed lines and the spatial location of magnetic O-type nulls are
shown as vertical red lines, X-type nulls as vertical blue dotted
lines. Right column: Wavelet power spectra of each of the corresponding
line plot quantities.  Animations of this figure are available as
electronic supplements to the online version.  \label{f6}
}
\end{figure*}


The left column of Figure~\ref{f6} shows the scaled magnetic field
magnitude $(B^2/(8\pi))^{1/2}$ sampled along the CS arc fits for
BCS1, FCS1/BCS2, and FCS2 from top to bottom, respectively. The
vertical dashed gray lines indicate the estimates of the CS boundaries
$(x'_L, x'_R)$, the location of the O-type nulls are shown as
vertical red lines, and the location of X-type nulls are shown as
dotted blue lines.
The right column of Figure~\ref{f6} shows the rectified wavelet
power distribution at spatial scale size $X$ (in units of grid scale
$\delta$) as a function of position $x'$ along CS arc for each of
the corresponding line plots for each CS.

The largest magnetic islands are clearly visible as enhancements
in the magnetic field magnitude line plots and the locations of the
O-type nulls (the center of the magnetic islands) occur at the peaks
of these enhancements. For the well-resolved large islands, the
wavelet transform power shows clear maxima at spatial scales
corresponding to the island size, $10\delta \lesssim X \lesssim
40\delta$.
The electronic animations of each CS in Figure~\ref{f6} ({\tt
FIGURE12\_bcs1.mp4}, {\tt FIGURE12\_fcs1.mp4}, {\tt FIGURE12\_fcs2.mp4})
show the temporal evolution of the island formation and growth as
well as their propagation along the CS arc and ejection in the
reconnection exhaust -- both in the line plots and in the wavelet
power spectra.

\subsubsection{Magnetic Energy Density Power Spectra}

The global wavelet power spectrum, or the integrated power per scale
(IPPS) for the Morlet family of wavelet transforms is roughly
equivalent to the global Fourier transform \citep{Le2003, Bolzan2005}.
Here we utilize the wavelet transform's spatial dependence to
construct the IPPS spectra of the magnetic energy density along the
CS arc in each timeframe only between the boundaries of the CS.
While there is still significant structure in the plasmoid quantities
just outside of these boundaries, the interaction of the magnetic
islands (e.g., reconnection, deformation, etc) with either the
line-tied flux system to the left or the CME and/or open flux to
the right represent a consequence of the CS evolution rather than
an intrinsic part of the reconnection dynamics in the CS itself.

\begin{figure}[t]
\center \includegraphics[width=20pc,height=4in]{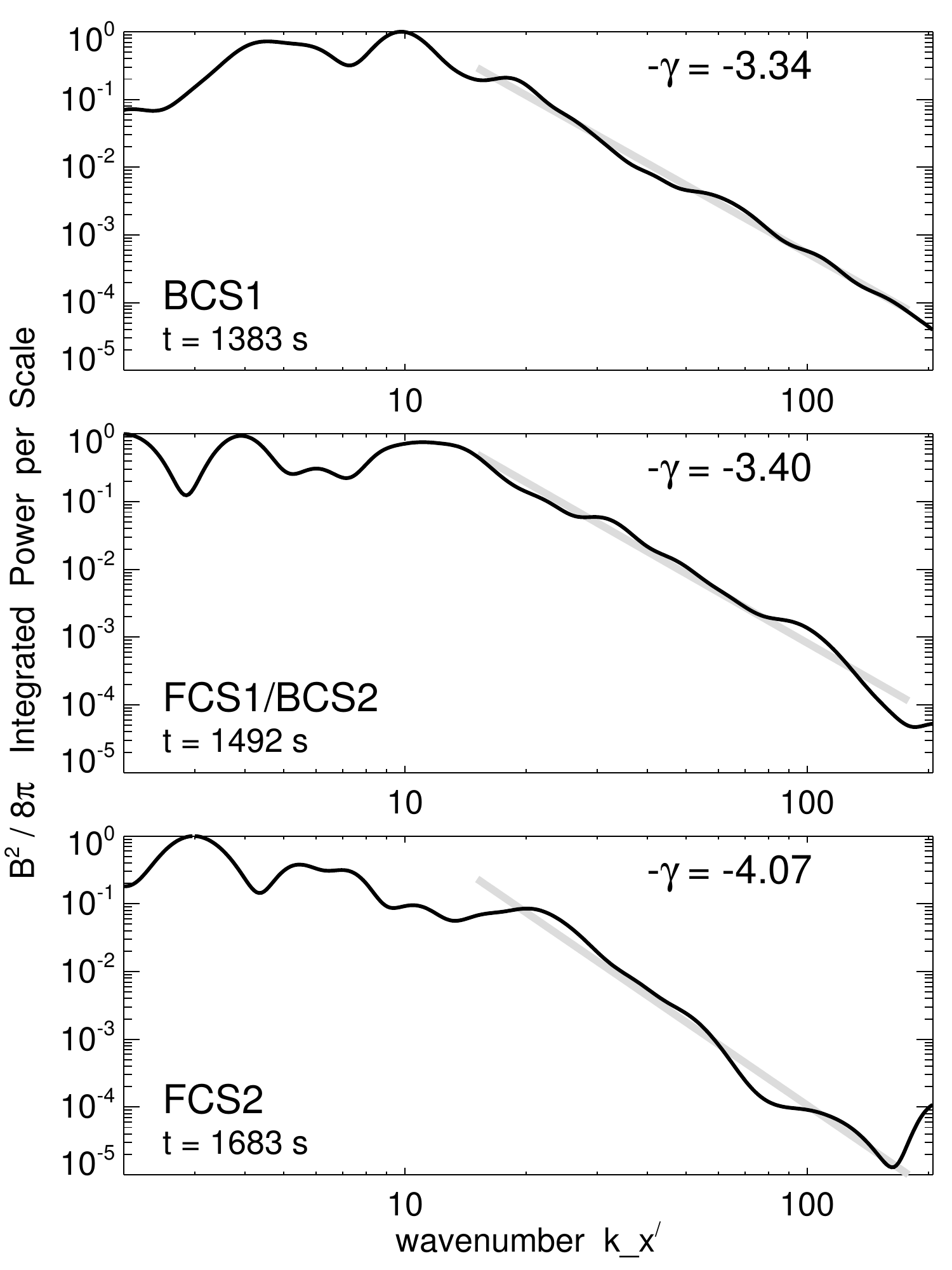}
\caption{Integrated power per scale (IPPS) spectra of the magnetic
energy density wavelet power for BCS1 (top), FCS1/BCS2 (middle),
and FCS2 (bottom). The power law fit to the spectra $k_{x'}^{-\gamma}$
for wave numbers $15 \le k_{x'} \le 180$ is shown as the thick gray
lines.  \label{f10} }
\end{figure}

Figure~\ref{f10} plots the IPPS spectrum for the wavelet power of
the magnetic energy density in each CS at the times shown in
Figure~\ref{f6}. We define a normalized spatial wavenumber $k_{x'}
= 4L_0 / X = 410\delta/X$ for ease of comparison with the standard
Fourier spectral analyses \citep[e.g.,][]{Shen2013}. Recalling
$\delta$ is the size of a single computational cell, the maximum
wave number of $k_{x'}=205$ corresponds to the Nyquist frequency
wavelet spatial scale of $X=2\delta$.
For the high frequency range of wavenumbers, $15 \le k_{x'} \le
180$, we fit the IPPS spectra with a power law of the form
$k_{x'}^{-\gamma}$. The power law fit is shown as the thick gray
line beneath the IPPS spectra curves for BCS1, FCS1/BCS2, and FCS2
spectra.

%
%
\begin{figure}[t]
\center \includegraphics[width=3.25in]{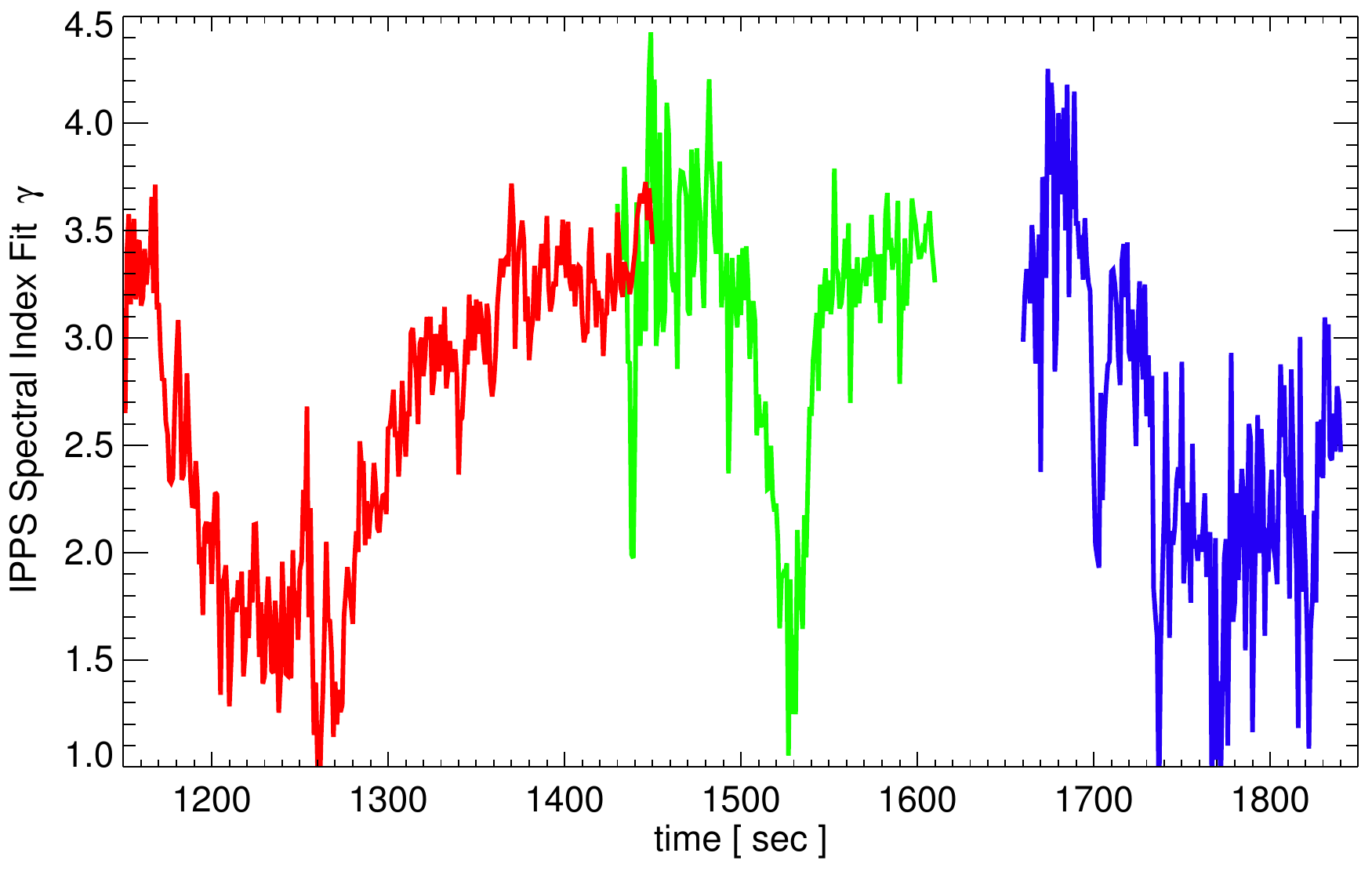}
\caption{Temporal evolution of the magnetic energy IPPS spectral
exponent $\gamma$ for each of our current sheets (BCS1 red; FCS1/BCS2
green; FCS2 blue). \label{f11} }
\end{figure}

Figure~\ref{f11} plots the temporal evolution of the spectral index
$\gamma$ in the usual color scheme (BCS1 red; FCS1/BCS2 green; FCS2
blue).
Overall, our results are in excellent agreement with those presented
by \citet{Shen2013}. We see agreement in both the range of the
magnetic energy density spectral index $1.5 \lesssim \gamma \lesssim
4$ and with their average value of $\gamma \sim 3.5$.
The temporal variability of $\gamma$ in our simulation shows moderate
fluctuations throughout the evolution of each of our sheets, with
periods of the largest variance associated with the rapid reconnection
phases in both FCS1/BCS2 and FCS2, but also a slower, more-extended
evolution that appears to reflect the global evolution of the CS
reconnection dynamics.

The IPPS exponent remains $\sim$2 during the initial development
of the X- and O-point chain in BCS1 over the period $1200 \lesssim
t \lesssim 1300$~s. While this early phase of the current sheet
elongation and onset of the plasmoid instability generate an
increasing number of islands, the size of these islands remain
relatively modest (cf. Figure~\ref{f12} showing the appearance of
only a couple islands with areas exceeding $10^2$~pixels through
$\sim$1300~s). However, from $t \gtrsim 1300$~s, $\gamma$ increases
to the value of $\sim$3.3
There is a noticeable, extended dip in the FCS1/BCS2 $\gamma$ from
$\sim$3.5 back to below 2 from about $1510 \lesssim t \lesssim
1550$~s even though this period does not have fewer islands (cf.
Figure~\ref{f2}) or corresponding changes in macroscopic sheet
properties (cf. Figure~\ref{f5a}). Inspection of the Figure~\ref{f3b}
movie (and also Figure~\ref{f12}) confirms that this period starts
with an ejection of a giant island ($A \sim 10^3$~pixels) and the
remaining islands are relatively small ($A \lesssim 40$~pixels) and
do not start to regularly exceed this size again until $t \gtrsim
1550$~s. This period also corresponds to the FCS1/BCS2 transition
between the eruptive flare reconnection of the first CME and the
overlying breakout reconnection preceding the second CME. The
guide-field component ($B_z$) is essentially zero here as it switches
signs and Figure~\ref{f15} shows the distribution of $B_z$ values
in the CS do not have the broadening to larger magnitudes over this
interval.
Finally, the FCS2 IPPS spectral index also shows the gradual
transition from $\gamma \sim 3.5$ down to $\sim$2, but in this case
there are corresponding changes in the macroscopic sheet properties,
i.e. the dissipation of the strong currents, shrinking of the CS
length, and the slowing down of the inflow and outflow speeds. For
$t \gtrsim 1740$, the island sizes remain small and the islands
themselves are very short-lived.
The qualitative behavior and evolution of our IPPS magnetic energy
density spectra appears entirely consistent with the physical
processes described by \citet{Shen2013} -- the spectra steepens
with island growth and merging and becomes shallower with the
ejection of the largest islands out of the sheet.

\section{Summary and Discussion}
\label{s:discussion}

%
%

We have presented a detailed analysis of the structure and evolution
of magnetic islands formed during reconnection in the three
large-scale, plasmoid-unstable current sheets associated with the
\citetalias{Lynch2013} sympathetic magnetic breakout eruption
scenario.
Our current sheets arise naturally and self-consistently from the
magnetic topology and evolution of a coronal psuedostreamer as a
response to the magnetic free energy introduced by gradual boundary
shearing flows and the subsequent rapid re-configuration of the
various flux systems during the initiation and eruption of sequential
CMEs.
The spatial and temporal resolution of the simulation is sufficient
to characterize the properties and dynamics of the onset and
development of the plasmoid instability in the overlying breakout
current sheet and both of the eruptive flare current sheets.

The intermittent, bursty emission that has been observed over a
wide range of wavelengths during solar flare events may be related
to the structure, dynamics, and evolution of magnetic islands in
eruptive flare current sheets \citep[e.g.,][and references
therein]{Kliem2000,Nakariakov2009}.
Pulses of enhanced radiation could originate in discrete acceleration
episodes associated with the formation and contraction of magnetic
islands during plasmoid-unstable reconnection.
In very high-resolution, adaptively-refined simulations of breakout
eruptive flare reconnection, \citet{Guidoni2016} have characterized
the contraction of different magnetic flux regions inside the MHD
simulation islands in order to estimate particle energy gain via
the \citet{Drake2006a} mechanism for electron acceleration in
plasmoid-unstable current sheets.

It is also important to highlight that in this particular \textit{ARMS}
simulation, our resistivity is entirely numerical. Our results and
analysis show that highly structured and detailed reconnection
dynamics can be obtained without an explicit, physical resistivity
term. The overall qualitative and quantitative properties of the
reconnection, i.e. the dimensionless reconnection rate, the magnetic
island size, mass, and flux content scaling, the magnetic energy
density spectral exponent, are comparable to results obtained via
resistive MHD codes, typically run with uniform resistivity.

An important next step in this arena of work will be the forward
modeling of synthetic observational signatures of plasmoid formation,
structure, and dynamics in the next generation of high-resolution
flare current sheet simulations. In particular, numerical MHD
simulations with a more realistic treatment of the energy equation
using field-aligned thermal conduction, ohmic dissipation, radiative
losses, and parameterized coronal heating, would allow for investigation
of the detailed thermodynamic evolution within the current sheet,
in the magnetic island plasmoids, and the interaction between
magnetic islands and flare arcade loops
\citep[e.g.,][]{Shen2013b,Downs2015}, and therefore enable a more
direct comparison to observations in the low corona.

\acknowledgments

The authors would like to thank the anonymous referee for valuable
suggestions during the review process and acknowledge Drs. George
Fisher, Spiro Antiochos, and Paul Cassak for helpful discussion
during the preparation of the manuscript.
B.J.L. and M.D.K. acknowledge support from AFOSR YIP FA9550-11-1-0048,
NASA HTP NNX11AJ65G, NSF AGS 1249150, and the Coronal Global
Evolutionary Model (CGEM) project NSF AGS 1321474. J.K.E. acknowledges
support from NASA LWS NNX10AQ616G. S.E.G. acknowledges support from
the NASA Postdoctoral Program at Goddard Space Flight Center,
administered by Oak Ridge Associated Universities.

\appendix

\section{Following the Current Sheets Through the Simulation Domain}
\label{S:Appendix1}

Our large scale current sheets are formed in response to the global
stresses and evolution of the magnetic field as an integral part
of the sympathetic CME eruption scenario from a pseudostreamer
topology. As the system evolves, our current sheets move through
the simulation domain. For the comparison between the properties
of the breakout and eruptive flare current sheets, we have presented
each in a similar rectangular $4L_0 \times 1.333 L_0$ region centered
on the current sheet. The rectangular regions are defined by three
time-dependent variables: the spatial coordinates of the rectangular
center $x_c(t)$, $y_c(t)$ and the rotation angle $\alpha(t)$ with
respect to the original domain's $x$-axis.
First, we estimate the position of the frame center and its orientation
angle by visual inspection of a subset of the images in the
Figure~\ref{f1} movie of $|J|$.
We then construct smooth, analytic functions of time based on our
initial visual inspection estimates. Thus, the time-evolution of
each of our current sheet-centered frames are given by
\begin{equation}
{\rm BCS1:} \;\;\; \begin{array}{lll}
		   x_c(t) & = & -0.0025 t + 2.8966 \\ 
		   y_c(t) & = &  0.00325 t + 1.7861 \\
		   \alpha(t) & = & 6.6 \tanh{\left[ \frac{t-1335}{50} \right]} + 55.4  
		  \end{array},
		  \label{eA1}
\end{equation}
for $t \in [1150, 1450]$~s, 
\begin{equation}
{\rm FCS1/BCS2:} \;\;\; \begin{array}{lll}
		   x_c(t) & = & 1.15 + 0.5\left( \frac{t-1430}{180} \right)^2 \\
		   	    &  & + 0.30\exp{\left[ -\left( \frac{t-1560}{30} \right)^2 \right]}\sin{\left[ 2\pi\left( \frac{t-1502}{50} \right)\right]} \\
		   y_c(t) & = &  4.85 - 3.5\left( \frac{t-1600}{180} \right)^2 \\
		   \alpha(t) & = & 105 - 20 \left( \frac{t-1610}{150} \right)^2  
		  \end{array},
		  \label{eA2}
\end{equation}
for $t \in [1430, 1610]$~s, and
\begin{equation}
{\rm FCS2:} \;\;\; \begin{array}{lll}
		   x_c(t) & = & 0.25 - 1.2\left( \frac{t-1830}{180} \right)^2 \\ 
		   y_c(t) & = &  0.875 \tanh{\left[ \frac{t-1680}{25} \right]} + 2.525 \\
		   \alpha(t) & = & 27.5 \tanh{\left[ \frac{1680-t}{50} \right]} + 72.5  
		  \end{array},
		  \label{eA3}
\end{equation}
for $t \in [1661, 1841]$~s. Here, $x_c$, $y_c$ are in units of $L_0$
and $\alpha$ is given in degrees.

The transformation from the original simulation coordinates $( x,
y )$ to the rectangular current sheet-centered coordinates $( x',
y' )$ are given by the standard rotation and translation formula
\begin{equation}
\left[ \begin{array}{c} x' \\ y' \end{array} \right] = 
\left[ \begin{array}{cc} \cos{\alpha(t)} & \sin{\alpha(t)} \\ \sin{\alpha(t)} & \cos{\alpha(t)} \end{array} \right]
\left[ \begin{array}{c} x-x_c(t) \\ y-y_c(t) \end{array} \right] .
\end{equation}

\begin{figure*}
\center \includegraphics[width=6.5in]{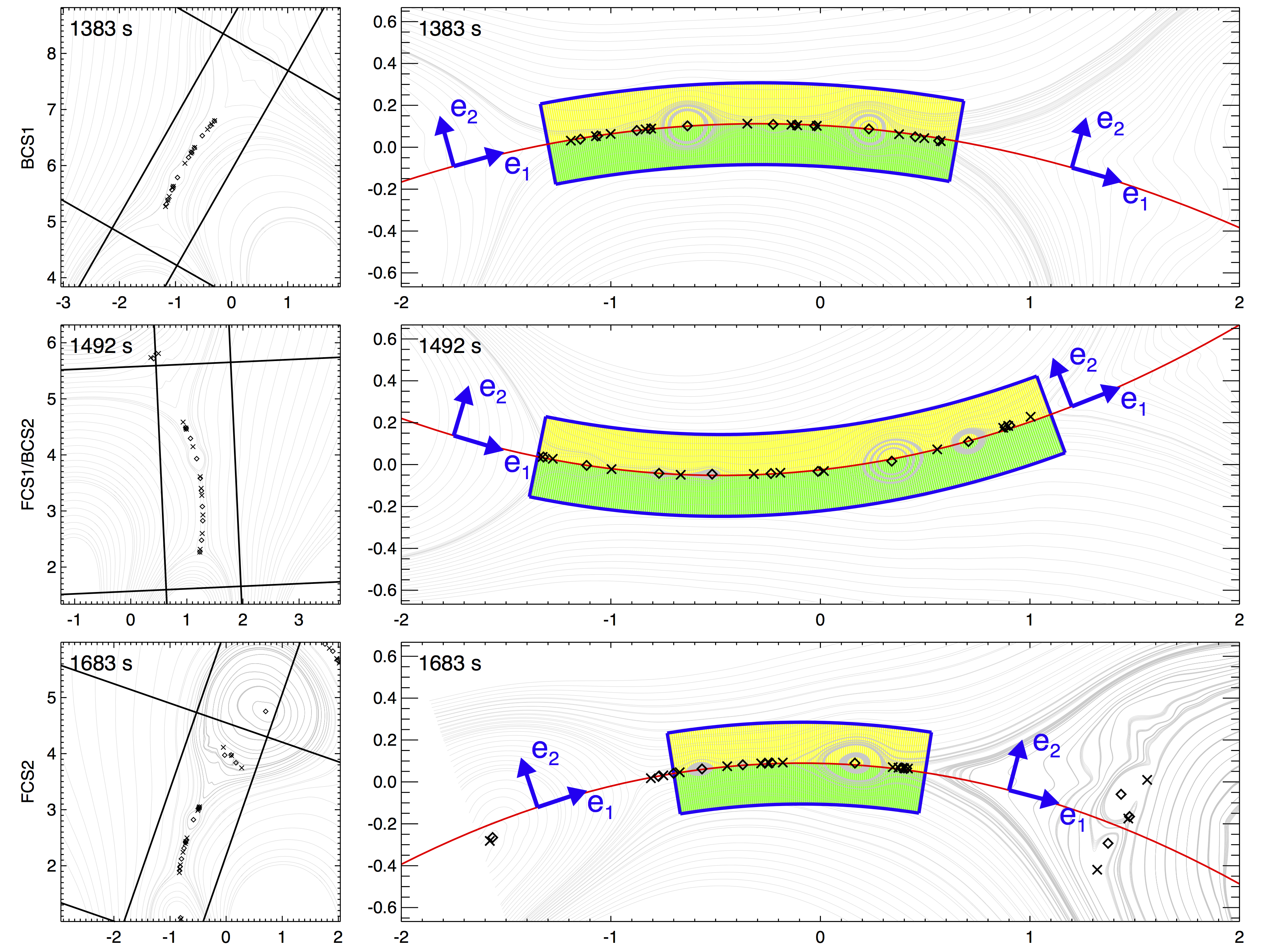}
\caption{Left column shows the location of the $(x', y')$ frames
centered on the three CS (BCS1 top; FCS1/BCS2 middle; FCS2 bottom)
described in Appendix~\ref{S:Appendix1}. The location of X-type
nulls (crosses) and O-type nulls (diamonds) are plotted in each
panel. Right column plots field lines in the $(x', y')$ frames along
with the curvilinear arc fits $f(x'(t_k))$ as the thin red lines
and the CS region boundaries as thick blue lines
(Appendix~\ref{S:Appendix2}). The \emph{local} current sheet
coordinates $(\mathbf{\hat{e}_1}, \mathbf{\hat{e}_2}$) are indicated
with unit vectors at representative positions along the curvilinear
arc fits in each frame. The upper and lower inflow region areas for
calculating CS-averaged quantities are shaded as yellow and green,
respectively (Appendix~\ref{S:Appendix3}). \label{f4} }
\end{figure*}

\noindent The left column of Figure~\ref{f4} plots the rectangular
regions centered on the three currents sheets: BCS1 ($t=1383$~s,
top row), FCS1/BCS2 ($t=1492$~s, middle row), and FCS2 ($t=1683$~s,
bottom row). The right column of Figure~\ref{f4} plots representative
field lines of the current sheet region in the $( x', y' )$ coordinate
frames.
Figures~\ref{f3a}, \ref{f3b}, and \ref{f3c} show the plasma properties
for each of the three current sheets from this perspective at these
time periods, respectively.
%

%
%

\section{Defining a Current Sheet-Centered Curvilinear Coordinate System}
\label{S:Appendix2}

While the $( x', y' )$ frames compensate for most of the bulk current
sheet motion, Figure~\ref{f4} also makes clear that the current
sheets have large-scale curvature. Thus, we use properties of the
current sheet to create a \emph{local, spatially-varying} curvilinear
coordinate system in order to calculate the sheet-averaged quantities,
specifically the corrected inflow and outflow velocities.
We fit a second-order polynomial to the spatial positions of the
null points to obtain the current sheet arc.
For the set of X- and O-type null points at location $\{ x'_i, y'_i
\}$, we fit a parabola of the form
\begin{equation}
f(x') = a x'^2 + b x' + c
\label{eb1}
\end{equation}
by minimizing the mean square error between the functional fit and
the null point locations
\begin{equation}
\chi^2 = \frac{1}{N}\sum_{i=0}^{N-1} \frac{ \left( f(x'_i) - y'_i \right)^2 }{\sigma_i} ,
\label{eb2}
\end{equation}
with weighting factors $\sigma_i$ based on spatial position, 
\begin{equation}
\sigma_i = \left\{ \begin{array}{ll}
		    0.05 & {\rm for} -1 \le x'_i \le +1 , \\ 
		    0.05 + 0.15\sin{\left( 0.5 \pi |x_i'-1.0| \right)} & {\rm otherwise} . \\ 
		   \end{array} \right.
\end{equation}
At times when there are less than 3 null points, we take the CS arc fit to be a horizontal line at the mean $y'$ value.

We apply a second, ``corrector" step based on the initial $f(x')$
fit. At 50 intervals evenly spaced in $x'$ we sample $|J|$ in $\pm$10
grid cells in $y'$ centered on $f(x')$. The spatial locations of
$\max{\{|J(x',y')|\}}$ are then used in the least squared fit of
the form of Equations~(\ref{eb1}) and (\ref{eb2}) with weightings
based on the ratio between the current density magnitude at $x_i$
and the maximum $|J|$ over the whole set of $x_i$ samples:
$\sigma_i^{|J|} = 0.02\max{\{|J|\}}/|J(x'_i)|$.
This allows regions of strong current density in the vicinity of
the initial $f(x')$ estimate to exert some influence over the arc
fit when there are few nearby null points. For example, in the early
development of the flare current sheets FCS1/BCS2, FCS2, there is
a single X-point but still a well defined current sheet arc.

This procedure works well for the vast majority of the 660 simulation
frames analyzed here (300 for BCS1 and 180 each for FCS1/BCS2 and
FCS2).
However, for some frames, the above fitting procedure clearly misses
a portion of the current sheet. In these frames ($t \in \{1465,
1471, 1478, 1678, 1748, 1749, 1760, 1761, 1775, 1786 \}$~s) we
impose the fit arc parameters by averaging the good fits in the
adjacent frames.
Occasionally, the $|J|$ corrector step does not improve the arc fit
so we keep either the imposed or the original $f(x')$ parameters
for the following times:
$t \in \{1440,1442,1445,1465,1471,1478,1522,1551,1554-1563,1583,1584,1588-1597,1678,1698,1748,1749,1752,1775,1786\}$~s.

As current sheet evolves, the curve defining the CS spatial extent
also evolves. Every simulation output time $t_k$ has its own parabolic
arc $f_k = f(x'(t_k))$ fit to the CS. Each $f_k$ defines an
instantaneous, \emph{local} curvilinear coordinate system which can
be described by unit vectors $(\mathbf{\hat{e}}_1, \mathbf{\hat{e}}_2)$
where $\mathbf{\hat{e}}_1(x',y')$ is tangent to the $f_k(x')$ curve,
$\mathbf{\hat{e}}_1 \cdot \mathbf{\hat{e}}_2 = 0$, and $\mathbf{\hat{e}}_1
\times \mathbf{\hat{e}}_2 = \mathbf{\hat{z}}$:
\begin{eqnarray}
\mathbf{\hat{e}_1}(x',y') & = & \frac{1}{\sqrt{1+\left( \frac{df_k}{dx'} \right)^2}} \left( \mathbf{\hat{x}'} + \frac{df_k}{dx'} \mathbf{\hat{y}'} \right) , \label{eB4}\\
\mathbf{\hat{e}_2}(x',y') & = & \frac{1}{\sqrt{1+\left( \frac{df_k}{dx'} \right)^2}} \left( -\frac{df_k}{dx'} \mathbf{\hat{x}'} + \mathbf{\hat{y}'} \right) . \label{eB5}
\end{eqnarray}
The right column of Figure~\ref{f4} plots the location of the X-type
and O-type nulls (as crosses and diamonds), the $f_k(x')$ arc fit
as the red line, and $(\mathbf{\hat{e}}_1, \mathbf{\hat{e}}_2)$
unit vectors at two representative positions along $f_k(x')$ to
illustrate their spatial dependence.
The boundary of the CS region is highlighted as thick blue lines,
and each of the inflow regions shaded as yellow and green.

The total current sheet length $2L(t)$ is obtained via standard arc
length integration between our estimates of the CS boundaries,
$x'_L$ to $x'_R$, at each simulation output frame $t_k$,
\begin{equation}
2 L(t_k) = \int_{x'_L(t_k)}^{x'_R(t_k)} dx' \sqrt{ 1 + \left( \frac{df_k}{dx'} \right)^2 } .
\end{equation}
The half-length $L(t_k)$ is used in the calculation of the CS aspect
ratio $L(t)/\delta$ shown the top panel of Figure~\ref{f5a}.  The
($x'_L$, $x'_R$) current sheet boundaries were obtained via visual
inspection of every other simulation time output frame ($t_k$ even)
and the positions during $t_k$ odd simulation times were linearly
interpolated between the positions of the adjacent even times. The
estimate of the CS boundaries were guided by the opening angle of
the field lines made with respect to the parabolic arc fit.

\section{Constructing Current Sheet-Averaged Quantities}
\label{S:Appendix3}

The \emph{local, spatially-varying} CS curvilinear coordinates
defined by Equations~\ref{eB4}, \ref{eB5} are used to decompose the
velocity and magnetic field vectors into components tangent to
($\mathbf{\hat{e}}_1$) and perpendicular to ($\mathbf{\hat{e}}_2$)
the current sheet.
The precise inflow and outflow velocities are given by $V_{in}^{\pm}
= \pm (\mathbf{V} \cdot \mathbf{\hat{e}}_2)$ and $V_{out}^{\pm} =
\pm (\mathbf{V} \cdot \mathbf{\hat{e}}_1)$, where the $\pm$ notation
indicates the positive, negative $\mathbf{\hat{e}}_1$, $\mathbf{\hat{e}}_2$
directions respectively: i.e., inflow into the sheet from `above'
$(-)$ and `below' $(+)$, and outflow from the sheet to the `left'
$(-)$ and `right' $(+)$.

To obtain the sheet-averaged quantities, we construct the mean
`above' and `below' inflow components $\langle V_{in}^{-} \rangle$,
$\langle V_{in}^{+} \rangle$ by averaging the $\mathbf{V} \cdot
\mathbf{\hat{e}}_2$ values in the yellow and green regions shown
in Figure~\ref{f4}.
The inflow regions are defined by a distance of 20 grid points in
the direction of $\pm \mathbf{\hat{e}}_2$ starting from $(x', f(x'))$
over the range of the CS boundaries $x' \in [x'_L, x'_R]$.
We also construct the mean `left' and `right' outflow velocities,
$\langle V_{out}^{-} \rangle$ and $\langle V_{out}^{+} \rangle$ in
an analogous fashion from the $\mathbf{V} \cdot \mathbf{\hat{e}}_1$
values. Here we take a single line of $\pm$5 grid points in the
$\mathbf{\hat{e}}_2$ direction centered on $(x'_L, f(x'_L))$ for
the `left' boundary and $(x'_R, f(x'_R))$ for the `right' boundary.
In both the inflow and outflow cases, the $(+)$ and $(-)$ values
have the opposite sign, so we average the two sets of magnitudes
to get the CS-averaged quantities:

\begin{eqnarray}
   \langle V_{in} \rangle & = & \onehalf \left(  \langle V_{in}^{+} \rangle  +  \langle V_{in}^{-} \rangle  \right) ,\\
   \langle V_{out} \rangle &=& \onehalf \left(  \langle V_{out}^{+} \rangle  +  \langle V_{out}^{-} \rangle  \right) .
\end{eqnarray}
These quantities are plotted for each of the current sheets in the
top panel of Figure~\ref{f5b} and their ratio $\langle V_{in} \rangle
/ \langle V_{out} \rangle$ is shown in the bottom panel of
Figure~\ref{f5a}.

The magnetic field can also be decomposed into components parallel
($\mathbf{B} \cdot \mathbf{\hat{e}}_1$) and perpendicular ($\mathbf{B}
\cdot \mathbf{\hat{e}}_2$) to the current sheet. We can estimate
the current sheet-averaged time rate of change in flux associated
with the inflow region by starting with the induction equation
(\ref{e4}), taking an area integral ($d\mathbf{A} = de_2 dz
\mathbf{\hat{e}}_1$), and applying Stokes' theorem:
\begin{eqnarray}
\int d\mathbf{A} \cdot \frac{\partial\mathbf{B}}{\partial t} & = &  \int d\mathbf{A} \cdot \nabla \times \left( \mathbf{V} \times \mathbf{B} \right) \nonumber \\
\frac{\partial}{\partial t} \int dA \left( \mathbf{B} \cdot \mathbf{\hat{e}}_1 \right) & = & \oint d \boldsymbol{\ell} \cdot \left( \mathbf{V} \times \mathbf{B} \right) .
\end{eqnarray}
Utilizing a convenient choice of area, the line integral can be
constructed to give the familiar result in terms of the $z$-component
of ($\mathbf{V} \times \mathbf{B}$):
\begin{equation}
\frac{d(\Phi_{in}/z) }{dt} = \left( \mathbf{V} \cdot \mathbf{\hat{e}}_1 \right)\left( \mathbf{B} \cdot \mathbf{\hat{e}}_2 \right) - 
						           \left( \mathbf{V} \cdot \mathbf{\hat{e}}_2 \right)\left( \mathbf{B} \cdot \mathbf{\hat{e}}_1 \right) .
\end{equation}
Likewise, the change in flux from the outflow is calculated with
the area ($d\mathbf{A} = de_1 dz \mathbf{\hat{e}}_2$) and its
corresponding line integral to obtain
\begin{equation}
\frac{d(\Phi_{out}/z) }{dt} = -\left( \mathbf{V} \cdot \mathbf{\hat{e}}_1 \right)\left( \mathbf{B} \cdot \mathbf{\hat{e}}_2 \right) + 
						          \left( \mathbf{V} \cdot \mathbf{\hat{e}}_2 \right)\left( \mathbf{B} \cdot \mathbf{\hat{e}}_1 \right) .
\end{equation}
We then apply the spatial averaging procedure described above for
the upper and lower portions of the inflow region to obtain $\langle
d(\Phi_{in}/z)/dt \rangle$
and along the left and right arc boundaries to obtain $\langle
d(\Phi_{out}/z)/dt \rangle$ (see middle panel of Figure~\ref{f5b}).
The total fluxes transferred through the sheet via inflow and outflow
are simply calculated as $\langle \Delta (\Phi/z) \rangle = \int
dt \; \langle d(\Phi/z)/dt \rangle$, shown in the bottom panel of
Figure~\ref{f5b}.




%

\clearpage

\end{document}